\documentclass[10pt,pra,superscriptaddress,amsmath,amssymb,twocolumn]{revtex4-1}
\pdfoutput=1

\usepackage{amsthm}
\usepackage{graphicx}
\usepackage[all]{xy}
\usepackage[usenames,dvipsnames]{color}
\usepackage{tabularx}

\newcommand{\mc}[1]{\mathcal #1}

\newcommand{\ave}[1]{\langle #1 \rangle}

\newcommand{\tr}{{\rm Tr}}
\newcommand{\one}{{\bf 1}}
\newcommand{\re}{{\rm Re}\,}
\newcommand{\im}{{\rm Im}\,}

\theoremstyle{plain}

\theoremstyle{definition}

\begin{document}

\title{Renormalisation as an inference problem}
\author{C\'edric B\'eny}
\author{Tobias J.\ Osborne}
\affiliation{Institut f\"ur Theoretische Physik, Leibniz Universit\"at Hannover, Appelstra{\ss}e 2, 30167 Hannover, Germany}
\date{\today}

\begin{abstract}
In physics we attempt to infer the rules governing a system given only the results of imprecise measurements. This is an ill-posed problem because certain features of the system's state cannot be resolved by the measurements. However, by ignoring the \emph{irrelevant} features, an \emph{effective theory} can be made for the remaining observable \emph{relevant} features. We explain how these relevant and irrelevant degrees of freedom can be concretely characterised using quantum distinguishability metrics, thus solving the ill-posed inference problem. This framework then allows us to provide an information-theoretic formulation of the renormalisation group, applicable to both statistical physics and quantum field theory. Using this formulation we show that, given a natural model for an experimentalist's spatial and field-strength measurement uncertainties, 
the $n$-point correlation functions of bounded momenta emerge as relevant observables.
Our methods also provide a way to extend renormalisation techniques to effective models which are not based on the usual quantum field formalism. In particular, we can explain in elementary terms, using the example of a simple classical system, some of the problems occurring in quantum field theory and their solution. 
\end{abstract}

\maketitle

\section{Introduction}

In the natural sciences we want to discover the rules that govern the natural world. The primary input for this task is quantitive data gathered from experiments. Thus we are continually confronted with the task of \emph{inferring} from noisy data a simple and economic explanation for the behaviour of complex interacting systems. 

At first sight, such a goal might seem hopelessly ambitious: 
even if there are simple unifying laws describing Planck-scale quantum gravitational physics, how could they manifest themselves in the conductivity of a metal or the motion of a tennis ball? The answer, of course, is that we can discover simple intermediate effective laws useful for the understanding of such large objects. The explanation of why and how such effective laws emerge, falling under the rubric of the \emph{renormalisation group} (RG), is one of the most profound ideas in physics.

The RG, as conceived by Wilson \cite{wilson:1974a,wilson:1975a}, shows why it is possible to describe long-distance physics while essentially ignoring short-distance phenomena; Wilson argued that, if we are content with predictions to some specified accuracy, the effects of physics at smaller lengthscales can be absorbed into the values of a few parameters of some \emph{effective (field) theory} for the long-distance degrees of freedom. This is the reason why physics at one lengthscale is effectively decoupled from physics at different length scales. 

The RG now underpins much of our understanding of modern theoretical physics and has been applied in a dazzling array of incarnations to study systems from quantum field theory to statistical physics \cite{fisher:1998a}, applied mathematics \cite{barenblatt:1996a}, and beyond. The central concept at the heart of this panoply is that, as information is lost, a theory valid for long-distance physics must \emph{flow} to a different \emph{simpler} theory. This observation cries out \cite{preskill:2000a} for a unifying information-theoretic formulation of the RG.

The task of developing an information theoretic framework for the RG has been attempted by several authors (see, e.g., \cite{machta13, apenko:2012a,brody:1998a,casini:2007a,gaite:1996a} for a selection), however, there are still several major remaining obstructions. The most fundamental problem is that there are actually two conceptually rather different versions of the RG, a ``quantum field-theoretic'' RG describing the flow of theories induced by changing an ultraviolet cutoff and a ``statistical physics'' RG describing the flow of theories resulting from zooming out from a fixed system. Wilson persuasively argued \cite{wilson:1975a}, in the path integral context, that these two RGs are actually equivalent. Unfortunately it is very difficult to imagine how to proceed with the path-integral framework if we want to build a purely information-theoretic formulation of the RG. While there are plenty of alternatives to the path integral incarnation, most notably, the Kadanoff block-spin RG \cite{kadanoff:1976a, kadanoff:1966a, kadanoff:1977a}, it is still very far from obvious how to apply it in an information theoretic way to explain the quantum field implementations of the RG. 

The objective of this paper is to develop a fully general and abstract information-theoretic framework for the RG, appropriate both for the QFT and statistical physics context. In pursuing this goal we found it necessary to first step back and reconsider the information-theoretic task of \emph{inference} in quantum mechanics. We begin by phrasing this task as a game played between two players: Alice, who possesses a quantum system, and Bob, who perceives the system via a noisy quantum channel. When Bob tries to infer the state of Alice's systems, he is faced with the ill-posed \emph{inverse problem} of inverting a quantum channel to find the input from the output. This task is not well-posed because there exist \emph{equivalence classes} of states which lead to the same output of the channel. We discuss the optimal solution to this inverse problem by introducing the concept of \emph{relevance} which allows us to quantify what features of a quantum state are important for the solution of the inverse problem. By exploiting certain \emph{eigenrelevance} operators we then stabilise the inversion task rendering it well posed: we argue that a smooth and unique parametrisation of the equivalence classes is possible.  With the inference task now solved we then discuss both formulations of the RG within a common framework: it is argued that, in both cases, the RG gives a flow on an equivalence class of indistinguishable states. We conclude the paper by applying this framework to a variety of examples both from classical and quantum physics.

There are several dividends paid by this investment in a general information-theoretic formulation:
\begin{enumerate}
\item The equivalence classes induced by the channel modelling Bob's observational limitations allow us to give a precise definition for what is meant by \emph{effective state} and, correspondingly, \emph{effective theory}.
\item The information-theoretic framework developed also allows us to give explanations, in very simple terms, of some of the phenomena present in discussions of the QFT RG, including, divergences, regularisation, and renormalisability.
\item We resolve an issue noticed by Wilson \cite{wilson:1975a}: in the usual QFT setting, the eigenvalue equation determining the relevant eigenoperators near a fixed point does not come from a hermitian operator. By exploiting information metrics we always obtain a hermitian operator for the eigenoperators. 
\item We present a general channel which models Bob's limitations in the case where his spatial resolution is finite and then compute the eigenrelevance operators in a wide variety of settings, including, for small quantum systems, classical single-particle systems, classical field theories, quantum systems with continuous degrees of freedom, and quantum field theories. 
\item These calculations establish the central role played by the $n$-point correlation functions in QFT: these correspond to the eigenrelevance observables when Bob's ability to resolve local degrees of freedom is limited. 
\item A further consequence is an explanation for why Gaussian theories emerge as good effective theories, because the two-point correlation functions turn out to be the most relevant observables.
\item Finally, we clear up a little mystery present in many discussions of the QFT RG: why, when information is being lost, does one speak of a pure state for the system? The resolution is now simple: as we are dealing with the task of inferring the \emph{input} to a channel there is no reason the solution needs to be mixed. 
\end{enumerate}

\section{Overview}

\subsection{Inference}
Many tasks in physics can be summarised as the attempt to understand the state of a system given only limited experimental data.
Suppose that Alice (mother Nature) possesses this system and Bob is the experimentalist. Bob's task is to build a model of Alice's system $A$ which reproduces all the experimental results he has so far obtained. Because he has limited resources his experimental apparatus can only measure certain observables of the form $\mathcal{E}^\dag(M)$ \footnote{Note here that $M$ is a POVM element and $\mathcal{E}^\dag(M)$ represents the result of one of Bob's yes/no measurements applied to $A$. We do not assume that Bob can measure all observables of the form $\mathcal{E}^\dag(A)$; in particular, Bob is \emph{not} assumed to be able to measure the projections  $P_j$ occurring in the spectral decomposition $\mathcal{E}^\dag(B) = \sum_{j} a_j P_j$. Rather, by measuring $B= \sum_{j}b_j Q_j$ on his system, Bob effectively measures the POVM with elements $\mc E^\dagger(Q_j)$ on Alice'es system.
}, where $\mathcal{E}$ is a completely positive map from $A$ to $B$:
\begin{center}
\includegraphics{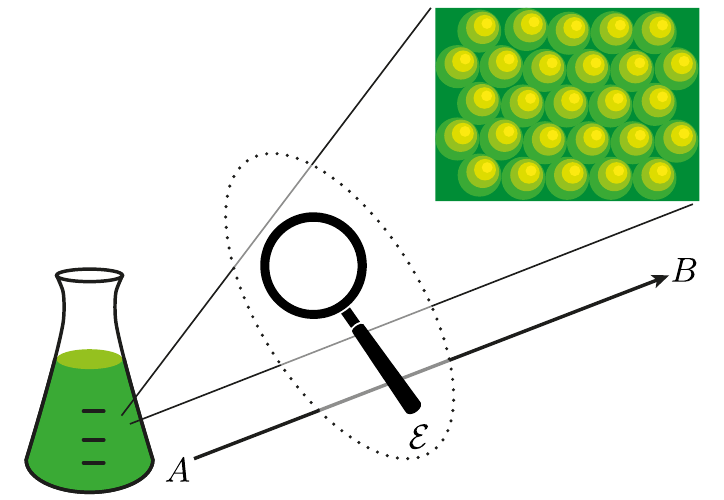}
\end{center} 
If Alice's system is in the state $\rho_{\text{true}}$ then we can summarise the information Bob can access using his apparatus with the state $\rho_B = \mathcal{E}(\rho_{\text{true}})$:
\begin{center}
\includegraphics{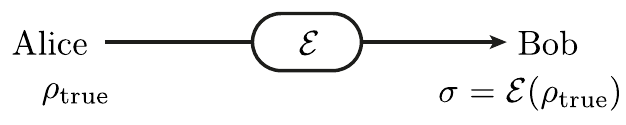}
\end{center}
(There is no need for  Alice's Hilbert space $\mathcal{H}_A$ to be the same as Bob's Hilbert space $\mathcal{H}_B$.) 

Repeated experiments can be thought of as Alice sending Bob identical copies of her state $\rho_{\text{true}}$ through the channel $\mathcal{E}$, one after another; Bob's goal is to figure out as much as possible about $\rho_{\text{true}}$. For concreteness, we need to assume that Bob knows what $\mc E$ is and, therefore, also what Alice's Hilbert space $\mathcal{H}_A$ is, so that all the parameters left to be determined experimentally are encoded in $\rho_{\text{true}}$.
The channel $\mc E$ can be used to encode any type of experimental limitation, such as a finite ability to resolve lengths or energies. 

If Alice sends an infinite number of copies of $\rho_{\text{true}}$ then, in the generic case where $\mc E$ is invertible as a linear map, Bob may be able to do full tomography of the state $\rho_B = \mc E(\rho_{\text{true}})$ and compute the density matrix $\rho_{\text{true}} = \mc E^{-1}(\rho_B)$. However, since the number of copies at Bob's disposal is always finite, he is left with some uncertainty about the exact values of the matrix elements of $\rho_B$, and hence $\rho_{\text{true}}$. This is a serious problem if $\mc E$ decreases the distinguishability between orthogonal pairs of states beyond Bob's tomographic abilities because he is left with an ill-conditioned \emph{inverse problem} which is unstable and usually does not have a unique solution.  
This is the generic situation in fundamental physics and there is no way to deal with it without extra assumptions. 

The appearance of an inverse problem does not deter Bob and his colleagues because all he really needs to proceed is a reasonable \emph{hypothesis} --- or \emph{effective state} --- $\rho$ which is indistinguishable from Alice's state with the current experimental limitations. 
With this hypothesis in hand experiments can be carried out to reject all competing hypotheses. If the hypothesis $\rho$ remains consistent with new experimental data as it comes in then the confidence that $\rho$ is a good explanation for Alice's state increases.
To quantify these statements we need discuss what ``indistinguishable'' means: we need to agree upon a measure of distance between quantum states.

As an example we use the relative entropy, whose operational interpretation is as follows.
Suppose there is a reigning orthodoxy amongst Bob's colleagues that Alice's state is $\rho$, but that Bob is trying to convince them that it is actually $\rho'$ instead. In this case the \emph{relative entropy} $S(\rho' \| \rho) \equiv \tr(\rho'(\log(\rho')-\log(\rho))$ is the natural measure of distinguishability to use. This quantity is exactly the optimal rate (per experiment) at which the (log of) the probability he mistakes $\rho'$ for $\rho$ decreases, while keeping the probability of making the opposite error small but constant~\footnote{This operational interpretation assumes that Bob is actually able to make joint quantum measurements on all his copies, which may be overly optimistic in general. A more appropriate quantity might be based on the task of characterising distinguishability using only LOCC measurements. As will become evident, however, our framework is easily applied to arbitrary information metrics.}. Here Bob's colleagues are demanding results with the highest level of confidence before they change their minds about what they consider to be the more surprising outcome.

An {\em effective state} $\rho$ is therefore one such that $\mc E(\rho)$ is approximately indistinguishable from $\mc E(\rho_{\rm true})$ according to $S(\cdot \|\cdot)$ with the current experimental limitations~\footnote{We observe that, even though $\mc E(\rho)$ may always be very mixed, the effective state $\rho$ may perfectly well be taken to be pure.}. 
This notion of indistinguishability suggests a notion of approximate equivalence \footnote{This notion of approximate equivalence does \emph{not} give us an equivalence relation at this stage because it is neither reflexive nor transitive.} between states: we say that states $\rho$ and $\rho'$ are \emph{approximately equivalent} from the point of view of Bob if he cannot distinguish them experimentally, i.e., if 
\[
S(\mc E(\rho)\|\mc E(\rho')) \le \epsilon,
\]
where the value of $\epsilon$ depends on the number of experiments he can afford to do, and on the confidence level he requires. 

\begin{figure}
\includegraphics[width=0.9\columnwidth]{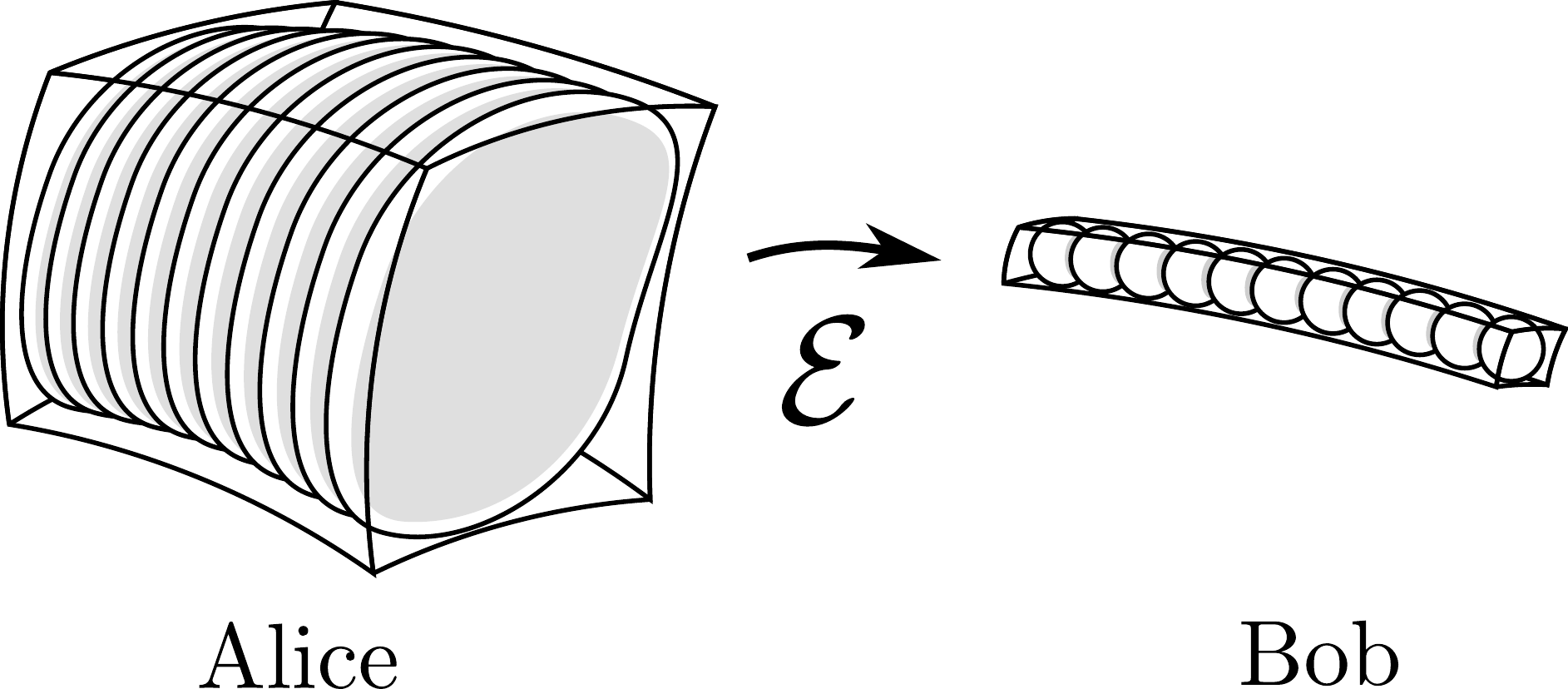}
\caption{Infinitesimal balls of approximately indistiguishable states on Bob's system typically correspond to large flat shapes on Alice's manifold.}
\label{pancakes}
\end{figure}

The set of states which are approximately equivalent to some state $\rho$ is the preimage of a small ball of states around $\rho_B= \mc E(\rho)$ under $\mc E$ (small in the sense that these states are very close to $\rho_B$ as measured using the relative entropy). We expect that the channel $\mc E$ greatly reduces the distinguishability of states along certain directions in the set of states~\cite{machta13}, which means that our sets of approximately equivalent states correspond, at least locally, to large pancake-like shapes on Alice's system (Fig.~\ref{pancakes}). 

\begin{figure}
\begin{tabularx}{0.9\columnwidth}{XllX}
(a) & (b) \\
\includegraphics[width=0.4\columnwidth]{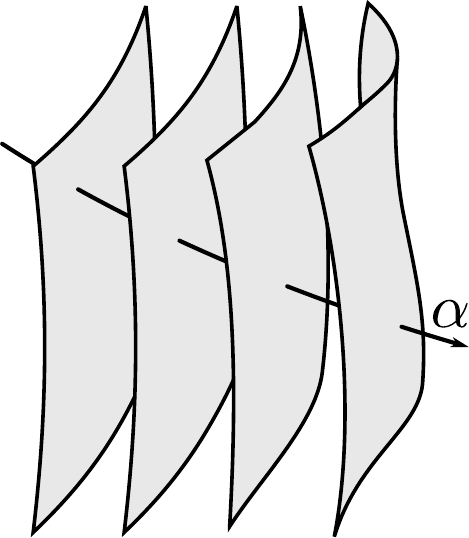} &
\includegraphics[width=0.4\columnwidth]{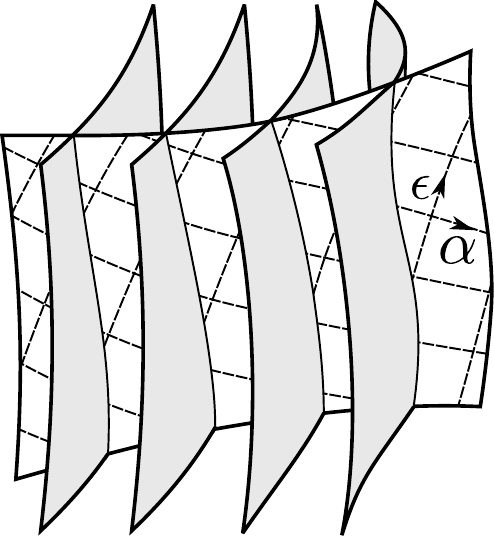} \\
\end{tabularx}
\caption{Schematic diagram of the three-dimensional state space of a fictitious system, with and without regularisation. The shaded planes represent equivalence classes of states which cannot be distinguished experimentally. They are intersected by the manifold of effective theories, parametrized in example (a) by a sole parameter $\alpha$, and, in example (b), additionally by a regularisation parameter $\epsilon$. The intersection lines are the renormalisation trajectories $\alpha(\epsilon)$. 
}
\label{foliation}
\end{figure} 

This suggests that we could idealize these pancakes as a continuum of lower-dimentional sheets by neglecting the directions which do not contract under $\mc E$, hence foliating Alice's manifold into true equivalence classes of states which are effectively indistinguishable for Bob. 
A good class of effective states would then be a smooth parameterization of unique representative of these equivalence classes (Fig.~\ref{foliation}a). This solves the inverse problem by effectively removing the ill-conditioned coordinates. 

Unfortunately, these pancakes of equivalent states may be very complicated; the relative entropy $S(\cdot \|\cdot)$ is difficult to compute in practice.
However, we can simplify the problem by focussing on states which are only infinitesimally different from $\rho$. A physical justification for this simplification is that, after many experiments have already been performed, the ``gross'' or ``large-scale'' differences between $\rho$ and all possible neighbouring states $\rho'$ have already been firmly eliminated so Bob is essentially only left with the task of sorting out the finer details.

Thus, to determine the parameters, or coordinates, which Bob cannot easily distinguish we may at first study this task in a small neighbourhood of state space surrounding a given hypothesis $\rho$: the problem is reduced to studying states $\rho+\epsilon X$ close to $\rho$ and understanding which \emph{features} $X$ Bob can most easily spot \footnote{We assume, for simplicity, that $\rho$ has full rank so that all possible features $X$ are traceless Hermitian operators.}.
What we are doing here is linearising Alice's curved state space $\mathcal{S}_A$ --- according to the ``distance'' measure $S(\cdot\|\cdot)$ ---  around the point $\rho$ and producing a new linear space $T_\rho \mathcal{S}_A$ of features to model those of Alice's states which are infinitesimally close to $\rho$:
\begin{center}
\includegraphics{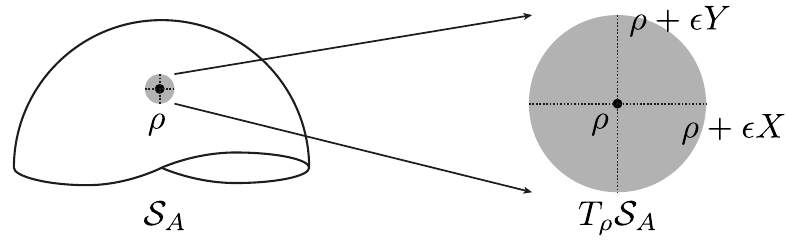}
\end{center} 
(The clumsy notation $T_\rho\mathcal{S}_A$ is inherited from its role as the \emph{tangent space} --- in the sense of differential geometry --- to the point $\rho$ in the manifold $\mathcal{S}_A$.)

We can  calculate the distance $S(\rho+\epsilon X\| \rho)$ from $\rho+\epsilon X$ to $\rho$ to lowest order in $\epsilon$: 
\[
S(\rho + \epsilon X \| \rho) = \epsilon^2 \tr(X \,\Omega_\rho^{-1}(X)) +\mc O(\epsilon^3),
\]
where the superoperator 
\[
\begin{split}
\Omega_\rho^{-1}(Y) 
& = \frac{d}{dt} \log (\rho + t Y) |_{t=0}
\end{split}
\]
is a non-commutative version of the operation ``division by $\rho$''~\footnote{This requires the observation that $\Omega_\rho^\dagger = \Omega_\rho$ and also that  $\Omega_\rho^{-1}(\rho) = \one$ for any $\rho$. Replacing $\rho$ with $\rho + t X$ and differentiating this last equation on both sides, we obtain $\frac{d}{dt} \Omega^{-1}_{\rho + t X}(\rho) = - \Omega^{-1}_\rho(X)$.}.
However, Bob can only perceive Alice's system via his experimental apparatus, which means that he can actually only measure the distinguishability between the states $\mathcal{E}(\rho+\epsilon X)$ and $\mathcal{E}(\rho)$:
\begin{equation}
	S(\mathcal{E}(\rho+\epsilon X) \| \mathcal{E}(\rho) ) =  \epsilon^2 \tr(\mathcal{E}(X) \,\Omega_{\mathcal{E}(\rho)}^{-1}(\mathcal{E}(X))) +\mc O(\epsilon^3).
\end{equation}
This quantity enjoys the same operational interpretation as for $S(\cdot  \| \cdot)$, but for an observer who can only effectively measure POVM elements of the form $\mc E^\dagger(M)$, which is precisely the situation Bob finds himself in relation to Alice's system.

Bob's reduced ability to distinguish $\rho + \epsilon X$ from $\rho$ is quantified by the ratio
\begin{equation}\label{eq:relevance}
	\eta_\rho(X) := \frac{ \ave{\mc E(X),\mc E(X)}_{\mc E(\rho)}}{\ave{X,X}_\rho},
\end{equation}
where $\ave{X,Y}_\rho \equiv \tr(X \,\Omega_\rho^{-1}(Y))$, which measures the statistical visibility of the state $\rho+\epsilon X$. We call this quantity the {\em relevance} of the direction $X$. The quantity $\ave{X,Y}_\rho$ is an inner product on the space $T_\rho\mathcal{S}_A$ of features/operators and allows us to measure not only the ``length'' or ``size'' of a feature, but also the ``angle'' between two features $X$ and $Y$ --- it is a \emph{metric} in the sense of differential geometry and is one of the many quantum generalizations of the Fisher information metric \cite{petz:1996a}.
The ratio Eq.~(\ref{eq:relevance}) crucially allows Bob to rank all the possible features $X$ according to their relevance: the smaller the value of $\eta_{\rho}(X)$ the less visible $X$ will be. 

A very simple example to keep in mind is the partial trace channel: suppose Alice's system is comprised of two qubits $A_1A_2$ and Bob can only access qubit $A_1$. Thus $\mathcal{E}(\rho) \equiv \tr_{A_2}(\rho)$. Suppose Bob hypothesises that Alice's state is $\rho=\mathbb{I}\otimes \mathbb{I}/4$. Then Bob concludes that any feature of the form $X\otimes \mathbb{I}$ has relevance equal to $1$ and any feature of the form $X\otimes Y$, with $\tr(Y) = 0$ has relevance $0$.

Using Eq.~(\ref{eq:relevance}) Bob can now work out what the $n$ \emph{most relevant} features are by solving an optimisation problem: he maximises $\eta_\rho(X)$ over all $n$-dimensional subspaces of traceless hermitian operators $X_j$ (this is simply an application of Ky Fan's maximum principle \cite{bhatia:1997a}). This is equivalent to solving a generalised eigenvalue problem and the answer can be immediately written down: Bob obtains a list $X_n$ of features, or \emph{eigenrelevance features}, with corresponding \emph{eigenrelevance} $\eta_n$.  

Let's order the eigenrelevance operators $X_n$ in decreasing order of eigenrelevance $\eta_1 \ge \eta_2 \ge \cdots$. Because of his experimental limitations, there is an $n$ after which Bob doesn't feel confident in detecting the presence of the corresponding feature $X_j$: any operator in the span of the directions $X_j$ with $j \le n$ is {\em relevant} and any operator in the span of the rest is simply {\em irrelevant}. 

Given a truncated list $\{X_j\}_{j=1}^n$ of relevant features $X_j$ we can now define an actual
notion of \emph{equivalence} for Bob: we say two nearby states $\rho + \epsilon Y_1 + \mc O(\epsilon^2)$ and $\rho + \epsilon Y_2 + \mc O (\epsilon^2)$ are in the same equivalence class {\em to first order}, if the difference $Y_2-Y_1$ is irrelevant at $\rho$.
This can be tested by checking if $Y_1 - Y_2$ is orthogonal to all the relevant features, namely,
\begin{equation}\label{eq:tsequivreln}
\ave{Y_1-Y_2,X_j}_\rho = 0,  \quad \forall j\le n.
\end{equation}
One can check that this indeed induces an equivalence relation on $T_\rho\mathcal{S}_A$.

Another way to reformulate this condition is as follows. Define the operators $A_i = \Omega^{-1}_\rho(X_i)$ and call them {\em eigenrelevant observables}. We will see that they indeed qualify as observables because they are dual to features of states. The above conditions then say that the states $\rho + \epsilon Y_1$ and $\rho + \epsilon Y_2$ are equivalent if they share the same expectation values for all relevant observables, i.e., for all 
\begin{equation}
\label{relobs}
A = \Omega^{-1}_\rho(X),
\end{equation}
where $X$ is a relevant feature.

We can illustrate this as follows. A small ball on Bob's system, containing all states whose distinguishability from $\mc E(\rho)$ is less than $\epsilon$, appears as a larger ellipsoid on Alice's system:
\begin{center}
\includegraphics[width=0.65\columnwidth]{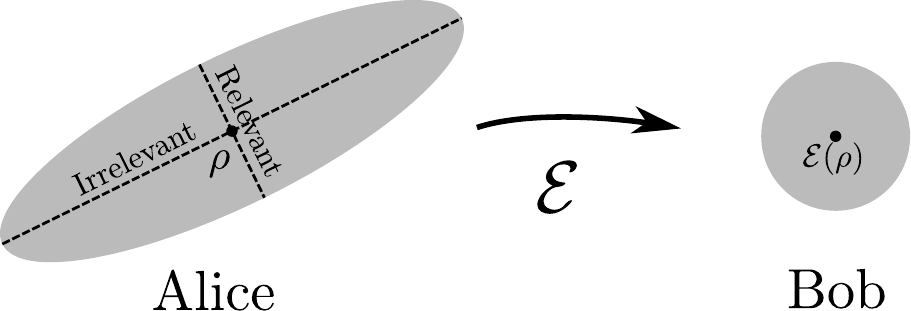}
\end{center}
The more stretched-out direction in this picture represent the {\em least} relevant features, because they contract the most under $\mc E$. Since all states in the ellipsoid are nearly indistinguishable for Bob, it constitutes our approximate equivalence class of states. 
The simplified equivalence class defined via Eq.~(\ref{eq:tsequivreln}) amounts to idealising the ellipsoid in Alice's space as a lower-dimensional plane in $T_\rho\mathcal{S}_A$:
\begin{center}
\includegraphics[width=0.8\columnwidth]{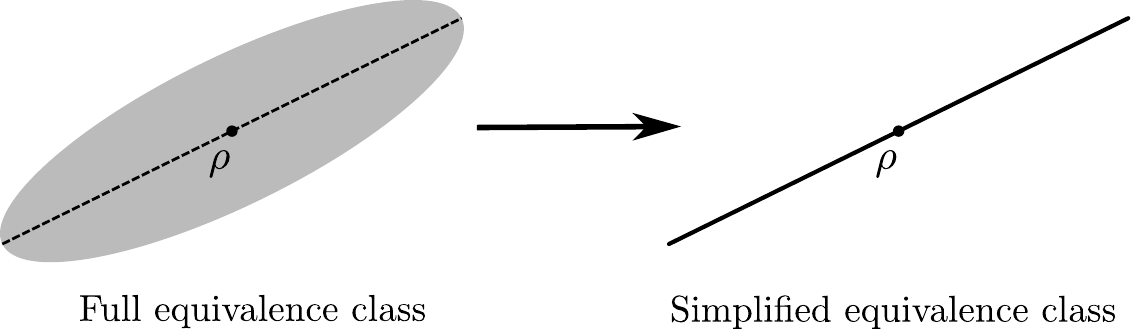}
\end{center}
The ellipsoid is simplified by sending the smaller principal axes (i.e., the more relevant directions) to zero, and the larger axes (i.e., the less relevant directions) to infinity.

The identification of these equivalence classes allows Bob to use a small set of effective states which only contain features that actually matter for the purpose of modelling his data.  A natural choice of effective states, to first order around $\rho$, is the family
\begin{equation}\label{expexp}
	\rho_\alpha = \rho + \sum_{j=1}^n \alpha_j X_j + \mc O(\alpha^2).
\end{equation}
Indeed, these states uniquely label the equivalence classes of states differing by a linear combination of irrelevant vectors, since they are linearly independent from the relevant ones. In addition, given a state $\rho + \epsilon Y + \mc O(\epsilon^2)$, Bob can determine the unique representative of its equivalence class, and hence solve his inverse problem, by projecting $Y$ onto the span of the relevant operators $X_j$, $j \le n$. The parameters of the corresponding effective state are simply
\[
\alpha_j = \epsilon \ave{X_j, Y}_\rho.
\] 
Although orthogonality with respect to the irrelevant directions is not essential for the purpose of representing the equivalence classes uniquely, it makes for the most rational model in the sense that it involves the minimal changes to the state needed to move from one equivalence class to the next (minimal as measured in both Bob's and Alice's metric).  

To summarise, we have decomposed the linear neighbourhood $T_\rho\mc S_A$ of each state $\rho$ into two orthogonal subspaces: the irrelevant directions $V_\rho \subseteq T_\rho\mc S_A$ and the relevant directions $V_\rho^\perp$. In this idealisation, states in the irrelevant neighbourhood are experimentally indistinguishable from $\rho$.

It is possible, although not necessarily true, that the irrelevant directions $V_\rho$ are tangent to some submanifold $\mc M_\rho$, i.e., such that $V_\rho = T_\rho \mc M_\rho$. If this is the case then the irrelevant fields can be {\em integrated} in order to find the manifold $\mc M$ passing through $\rho$. This submanifold could then serve as a reasonable definition for the nonperturbative equivalence classes of states containing $\rho$.  The same may be done for the orthogonal relevant fields, yielding a ``minimal'' effective manifold everywhere orthogonal to the irrelevant direction. We will see that the set of Gaussian states have this property for a reasonable choice of channel $\mc E$. 

This concludes the generalities for what Bob needs to do in order to build a model of Alice's system. To summarise: given the description of Bob the experimentalist's limited abilities, namely a channel $\mc E$, Alice's state space $\mathcal{S}_A$ may be foliated into equivalence classes of states which are approximately indistinguishable from Bob's point of view (for a given number of repetitions of the experiment). A good manifold of effective states is one which identifies a unique representant of each equivalence class (Fig.~\ref{foliation}a). This solves the ill-conditioned inverse problem of deducing the state from a coarse-grained measurement.

\subsection{The renormalisation group: statistical physics picture}
We are finally in a position to connect our framework with that of the renormalisation group. A challenging aspect of this objective is that a broad variety of concepts and methods fall under the rubric of ``renormalisation''. Following Wilson we roughly divide the renormalisation concept into two categories: (i) statistical physics renormalisation; and (ii) quantum field theoretic renormalisation. (We are certainly cognisant of the fact that this is perhaps too simplistic, but we believe it will be helpful for at least organising the reader's preconceived notions of the RG.) While these two categories appear, at least superficially, to be very different things, it was one of Wilson's great achievements to connect the two. In this subsection we'll explain the first category and in the following the second category. 

To discuss the RG in the context of statistical physics we must imagine that Alice has a possibly very complicated quantum system $A$. Bob can control this system by manipulating various \emph{external fields}, e.g., the pressure and the magnetic field. While Bob is pretty sure what Alice's hamiltonian $H_A$ is (i.e., he has worked out all the band structures and modelled the effects of all the interactions etc.) he is far from sure about the \emph{properties} of the Gibbs state $\rho_{\text{true}} = e^{-\beta H_A + \sum_j z_j A_j}/\mathcal{Z}$ as a function of the control field strengths $z_j$ because it is very difficult to exponentiate $H_A+\sum_j z_j A_j$. Bob gets around this by arguing that since his apparatus is insensitive to all the short-distance physics the only properties he can measure are long-distance degrees of freedom. Thus, since a lot of information is being lost, he should only really need to model large-scale collective degrees of freedom, i.e., his \emph{effective theory} of Alice's complicated system should be much \emph{simpler} than the exact model. (It is in this sense that thermodynamics can be understood as the ultimate effective theory --- this is the theory that emerges when all spatial information is neglected.)

This ``statistical physics'' picture fits into the previously described framework as follows: the span of the most relevant eigenrelevance operators $X_j$ corresponds to these long-distance degrees of freedom. It is Bob's act of simplifying his effective theory for Alice's system by discarding information that is called \emph{renormalisation}.

In physics, particularly in the statistical physics context, there are often one or more tuneable parameters $\sigma_j$, $j=1, 2, \ldots$, which model the accuracy of an experiment. A good example to keep in mind is simply the sensitivity of a detector: the smaller $\sigma$ is,  the more sensitive the detector. Other parameters include, for example, the number of experiments performed, the quality of the fabrication, the energy of the impact particles, etc. 

Typically, however, there is one convenient dominant parameter upon which a majority of the sensitivity of the experiment depends. Let's idealise our situation and index the map connecting Alice to Bob with this single parameter: $\mathcal{E}_\sigma$. It may also be quite convenient (although by no means necessary) to assume that $\sigma$ can be adjusted continuously.

In general, the linear space of relevant features could change arbitrarily as a function of $\sigma$, however, if $\sigma$ is meant to represent a monotone loss of information, we expect that if an operator is irrelevant for a given $\sigma$, it is also irrelevant for any larger $\sigma$. It follows that the only effect of an increase in $\sigma$ is an increase in the dimension $n(\sigma)$ of the space of irrelevant features.

If this is the case then, given a good effective state $\rho$, there is a priori no reason to modify $\rho$ as $\sigma$ {\em increases}, as it still yields correct predictions for the now smaller set of relevant observables. However, Bob may want to use this opportunity to simplify his effective state. By properly removing the features of the states that became unobservable, Bob can make apparent those features which stay important. For instance, if $\sigma$ is a lengthscale, the simplified model may converge to one that only contains universal information about its thermodynamical phase.

A simple example of this procedure is analysed in detail in Section~\ref{toy}. Here Alice has a stochastic classical system consisting of a single real variable, e.g., the position $x$ of a particle. Hence the true state to be discovered by Bob is a probability distribution on $\mathbb R$: $x \mapsto \rho(x)$. Bob's experimental limitation consists of a finite precision $\sigma$ at which he can resolves the particle's position. This can be modeled by a channel $\mc E$---in this case a stochastic map since the system is classical---whose effect is a convolution of Alice's probability distribution with a Gaussian of width $\sigma$. 

Bob's initial hypothesis is a simple Gaussian distribution, which we think of as a thermal state $\rho(x) \propto e^{-H(x)}$ for the Hamiltonian $H(x) = \frac{x^2}{2\tau^2}$. Our eigenvalue equation can be solved for this system, yielding the Hermite polynomials as eigenrelevance observables with the polynomial of degree $n$ having relevance $(\tau/\sigma)^{2n}$ for $\sigma \gg \tau$. 

Since the first $n$ Hermite polynomials span all degree $n$ polynomials this means that two nearby states are equivalent from the point of view of Bob exactly when they have the same first $n$ moments, where $n$ is the threshold chosen by Bob. 

For instance, suppose that Bob's most detailed model for Alice's state is defined by the Hamiltonian $H_0(x) = \frac{x^2}{2\tau_0^2} + \lambda x^4$. In the case Bob can only measure the first two moments, i.e.\ $n=2$, the state $e^{-H_0(x)}/\mathcal{Z}_0$ is equivalent to the thermal state for the simpler effective Hamiltonian $H_1(x) = \frac{x^2}{2\tau_1^2}$. The new parameter $\tau_1$ is easily computed as the second moment of $\rho_0$, so that $\rho_0$ and $\rho_1$ indeed share the same first two moments.

This map from $H_0$ to $H_1$ is one step of the renormalisation group:  the Hamiltonian has been simplified by exploiting the freedom in moving the state within the equivalence class of states. This can also be interpreted as a dependance of the effective Hamiltonian on $\sigma$ if the threshold is defined in terms of a minimal relevance $\eta_0$. Indeed, $\sigma$ being such that $(\tau/\sigma)^n \ge \eta_0 > (\tau/\sigma)^{n+1}$ justifies using the threshold $n$. 

The fact that the simplification procedures in this example stops as $\sigma > \tau \eta_0$, as all states becomes equivalent, is an artefact of this simple model.

In addition, this renormalisation group consists of discrete steps because the eigenrelevance operators form a discrete set. In the context of an infinite lattice, or of a field, they may take on continuous labels and the renormalisation group can then depend continuously on a precision parameter $\sigma$. Such an example will be analysed in Section~\ref{classicalfield}.

\subsection{The renormalisation group: quantum field theory picture}

The renormalisation group is often discussed in the context of quantum field theory. Here there are some additional subtleties that entail not only cosmetic changes but also introduce new conceptual difficulties.

Let's first deal with \emph{regularisation}. In quantum field theory it is relatively easy to propose a hypothesis for Alice's state which doesn't make sense without a \emph{regulator} $\epsilon$ because, otherwise, it would give infinite predictions for in-principle physically meaningful quantities. Such hypotheses arise when extrapolating some characteristic of Alice's state, already observed to be true for a finite number of experimentally accessible degrees of freedom, to apply to an \emph{infinite} number of degrees of freedom.  

Because the regularisation parameter $\epsilon$ relates to a degree of freedom which is not observable by Bob, the resulting set of effective states does not uniquely label the equivalence classes of states. A change in $\epsilon$ can be compensated by a change in the state's parameters so as to stay within a given equivalence class. This dependance is the RG flow in quantum field theory (Fig.~\ref{foliation}b).

To give a very simple example of what can go wrong, we use again the toy model introduced in the previous section. Suppose that Bob works with the threshold $n=4$, and treats the parameter $\lambda$ perturbatively to first order:
\[
\rho'(x) = e^{-x^2/\tau^2 - \lambda x^4} \approx e^{-x^2/\tau^2}(1 - \lambda x^4).
\]
(Note that here the {\em feature} by which we perturbe the gaussian state is $X(x) = - \lambda \,e^{-x^2/\tau^2} x^4$).
Using this perturbative approach he may well measure $\lambda$ and find that a small {\em negative} value fits his data nicely. 
However, if he were to then believe that the resulting Hamiltonian $H'(x) = x^2/\tau^2 + \lambda x^4$ is the true state of Alice's system he is in for some trouble because the corresponding thermal state cannot be defined (this Hamiltonian is not bounded from below). 

However, since any state which shares the same first four moments would be indistinguishable for Bob, he has a lot of freedom to fix his theory. For example, he can add a {\em regularisation} term of the form $\epsilon x^6$ to the Hamiltonian, in which case the Hamiltonian is bounded from below and the state is well defined no matter how small $\epsilon$ is. Although this term changes the second and fourth moment of the state, this effect can be compensated by appropriately modifying the parameters $\tau$ and $\lambda$ to $\tau(\epsilon)$ and $\lambda(\epsilon)$. The dependance of the parameters of the effective Hamiltonian (i.e., the \emph{coupling constants} -- which are $\tau$ and $\lambda$ in this example) on the regularisation parameter $\epsilon$ is usually expressed in terms of its derivative with respect to $\log \Lambda$, where $\Lambda = 1/\epsilon$ is a maximal energy scale above which all fluctuations are neglected. In this case we obtain an equation involving the \emph{beta functions}: $\beta_i(\Lambda) = \Lambda \frac{d}{d\Lambda}\alpha_i(\Lambda)$.

This flow of the effective state as a function of a regularisation parameter $\epsilon$ has no \emph{a priori} relationship to the flow generated by varying the noise parameter $\sigma$ discussed in the previous section, apart from the fact that in both case they move within the same equivalence class of states. Although conceptually very different, Wilson persuasively argued that those two concepts of RG flow are actually equivalent in many situations relevant to quantum field theory and statistical physics when the regularisation parameter $\epsilon$ is a minimal lengthscale~\cite{wilson:1975a}. This will be discussed in Section~\ref{sec:wilson}.

Since a regulator is an arbitrary --- often very coarse --- cutoff, the regularised theory parametrized by $\Lambda$ is not expected to make correct predictions when probed above that energy scale. Therefore, a truly fundamental theory of physics should make sense when taking the limit $\Lambda \rightarrow \infty$ while staying on the experimentally determined equivalence class of states. For it to ``make sense'' the expectation values of all the observables which can be (at least in principle) physically measured should converge to a finite value. 

If this limit does not exist, then it may simply be that the chosen effective manifold does not contain the True Theory of Everything. To fix this, a larger part of the equivalence class can be explored by adding extra parameters to the model, essentially by regarding one or more previously arbitrary regularisation parameters as related to coupling constants of a bigger class of theories.

The resulting theory, however, cannot be used to make higher energy predictions until experiments have become powerful enough to measure the new parameters (hence lowering the theshold to make them relevant). If it turns out that infinitely many parameters spanning all relevance levels must be added, then the theory is deemed {\em non-renormalisable}. 

This used to be considered a problem because, no matter how good our experiments, one would never be able to measure all the parameters of the theory. However, this is only a problem if one wishes to attain the True Theory of Everything valid \emph{in principle} for all length scales. This is no problem at all for the more pragmatic goal of correctly modelling all possible experiment below a certain energy level, i.e.\ to contend with effective theories which are well-defined for any finite of value of $\Lambda$.

\section{General framework}
\subsection{Primal picture}

Recall that a key role in our discussion is played by the bilinear form
\begin{equation}
\label{metric}
\ave{X,Y}_\rho := \tr(X \,\Omega_\rho^{-1}(Y)).
\end{equation}
This is a quantum version of the \emph{Fisher information metric}.  Given that Bob can only access Alice's state via the channel $\mc E$ he effectively works with a different reduced distinguishability metric given by
\[
\ave{X,Y}^{\mc E}_\rho := \ave{\mc E(X),\mc E(Y)}_{\mc E(\rho)}.
\] 
A crucial property of the metric Eq.~(\ref{metric}) is that it contracts under the action of a channel, which means that $\ave{X,X}^{\mc E}_\rho \le \ave{X,X}_\rho$. As we discussed previously, Bob's reduced ability to distinguish $\rho + \epsilon X$ from $\rho$ is quantified by the ratio
\[
\eta_\rho(X) := \frac{ \ave{\mc E(X),\mc E(X)}_{\mc E(\rho)}}{\ave{X,X}_\rho},
\]
which we called the {\em relevance} of the direction $X$.  (Note that the relevance is the ratio of the original and coarse-grained {\em stiffness}, studied for classical models in Ref.~\cite{machta13}.)

The quantity $\eta$ is always smaller than $1$ and, although a value of $zero$ implies complete irrelevance, it is in practice often very small  for many of the features $X$ in the examples we later consider.

The adjoint $\mc R_\rho$ of $\mc E$ at $\rho$ is defined by~\cite{ohya04}
\[
\ave{\mc R_\rho (Y), X}_\rho = \ave{Y, \mc E(X)}_{\mc E(\rho)}.
\]
Explicitly, it is
\[
\mc R_\rho = \Omega_\rho \mc E^\dagger \Omega^{-1}_{\mc E(\rho)}.
\]
We can use it to write Bob's metric in term of Alice's:
\[
\ave{X,Y}^{\mc E}_\rho = \ave{X,\mc R_\rho (\mc E(Y))}_\rho.
\]
The eigenrelevence features $X_n$ of the map $\mc R_\rho \mc E$ are now found from the eigenvector equation
\begin{equation}\label{eq:eigenreleqn}
	\mc R_\rho \mc E(X_n) = \eta_n X_n,
\end{equation}
and are complete and orthogonal in Alice's metric at $\rho$. If we choose $X_n$ to be normalised  then we can easily compute the component of any vector $Y$ in the $X_n$ direction via $\alpha_n = \ave{Y,X_n}_\rho$. Note that the eigenrelevance equation Eq.~(\ref{eq:eigenreleqn}) is an eigenvector equation for a self-adjoint operator $\mc R_\rho \mc E$. This observation resolves an issue noticed by Wilson (p.\ 784 in \cite{wilson:1975a}); by adapting the scalar product to the information metric we can render the operator determining  the relevant operators hermitian and so always obtain a complete basis of eigenrelevance operators.

Since $\ave{Y,X_n}^{\mc E}_\rho = \eta_n \ave{Y,X_n}_\rho$, we can think of the effect of the Bob's limitation as a contraction of the component of $Y$ along $X_n$ by the eigenrelevance $\eta_n$.

\subsection{Dual picture}

In the examples considered below, the operators $A_n = \Omega_\rho^{-1}(X_n)$ actually turn out to be much simpler than the $X_n$'s. This amounts to working with \emph{observables} rather than states. Indeed, observables can be thought of as \emph{cotangent vectors} as they map states to expectation values; the metric $\Omega_\rho^{-1}$ can be used to map tangent to cotangent vectors. In addition, if we write Bob's hypothesis as the equilibrium state $\rho = e^{-H}/Z$ then, since $\Omega_\rho^{-1}$ is the derivative of the log we have, to first order in $\epsilon$, that
\[
\rho + \epsilon X_n \approx \frac 1 Z\, e^{-H + \epsilon A_n}.
\]
Note that the normalization factor $Z$ is unchanged because the requirement that tangent vectors satisfy $\tr(X_n) = 0$ translates to the requirement that $\tr(\rho A_n) = 0$.

This means that we can also think about the operators $A_n$ as perturbations to the Hamiltonian defining the corresponding equilibrium state. The eigenvalue equation for the $A_n$s is given by 
\begin{equation}
\label{eq:dualeigenreleqn}
\mc E^\dagger \mc R_\rho^\dagger (A_n) = \eta_n A_n
\end{equation}
and is essentially the Heisenberg picture version of the eigenvalue equation on states. 

Moreover, for observables $A$ which are not completely irrelevant, i.e.\ such that $\eta_\rho(A) > 0$,  the above equation implies that $A = \mc E^\dagger (B)$ for some operator $B$. Hence $A$ has the form of an observable that Bob can measure. 

The metric evaluated for two observables $A = \Omega_\rho^{-1}(X)$ and $B = \Omega_\rho^{-1}(Y)$ becomes $\ave{X,Y}_\rho = \tr(A \Omega_\rho(B))$. In the classical commuting case this is just the correlation between $A$ and $B$. More generally, this quantity is given by the second-order derivative of the free energy:
\[
\tr(A \,\Omega_{\rho} (B) ) = - \frac{\partial^2}{\partial \alpha \,\partial \beta}F(\alpha,\beta)|_{\alpha = \beta = 0},
\]
where
\[
F(\alpha,\beta) = -\log \tr\,e^{-H + \alpha A + \beta B}
\]
is the free energy functional.

Alternatively, explicitly introducing the inverse temperature $\tau$ in $\rho = e^{-\tau H}/Z$, it can be shown \cite{lieb73} that
\[
\Omega_\rho(B) = \int_0^1 ds\,\rho^{1-s} B \rho^{s} = \rho \frac 1 \tau \int_0^\tau  ds\, B_s,
\]
where $B_s := e^{-s H} B e^{s H}$ is the imaginary time translation of $B$. It follows that
\[
\tr(A \,\Omega_{\rho} (B) ) = \frac 1 \tau \int_0^\tau \tr( \rho A_0 B_s) ds.
\]
If $A$ and $B$ are field operators, this may be expressed in terms of the familiar imaginary time two-point correlation functions.

\subsection{First-order equivalence relation}

We want to neglect the changes in the state in a direction which contracts a lot under the action of the channel.  Let us order the eigenvectors of Equ. \ref{eq:eigenreleqn} in decreasing order of relevance, i.e., such that $1 \ge \eta_1 \ge \eta_2 \ge \dots \ge 0$. We pick some threshold $n$ and decide to neglect all directions in the span of the eigenvectors $X_i$ with $i > n$, which we call {\rm irrelevant}. 

What this means is that we consider that two states $\rho_1$ and $\rho_2$, in the neighbourhood of $\rho$, are {\em equivalent} for Bob if their difference $\rho_2 - \rho_1$ is irrelevant in the above sense:
\[
\rho_2 - \rho_1 \in {\rm span}\{X_{n+1},X_{n+2},\dots\}.
\]

This condition can be reformulated in a physically more transparent way using the dual Heisenberg picture. We call an observable $A$ {\em relevant} if it belongs to the span of the eigenvectors $A_1$, $A_2$, \dots, $A_n$ of Equ.~\ref{eq:dualeigenreleqn}, or, equivalently, if they are of the form $A = \Omega_\rho^{-1}(X)$ where $X$ is orthogonal to the linear space of irrelevant vectors. 

In terms of these observables, the two state $\rho_1$ and $\rho_2$ then are then considered equivalent if they yield the same expectation values for all relevant observables, i.e., if
\begin{equation}
\label{sameexpval}
\tr(\rho_1 A_i) = \tr(\rho_2 A_i) \quad \forall \, i \le n. 
\end{equation}

\subsection{Nonperturbative equivalence relation}

\begin{figure}
\includegraphics[width=0.8\columnwidth]{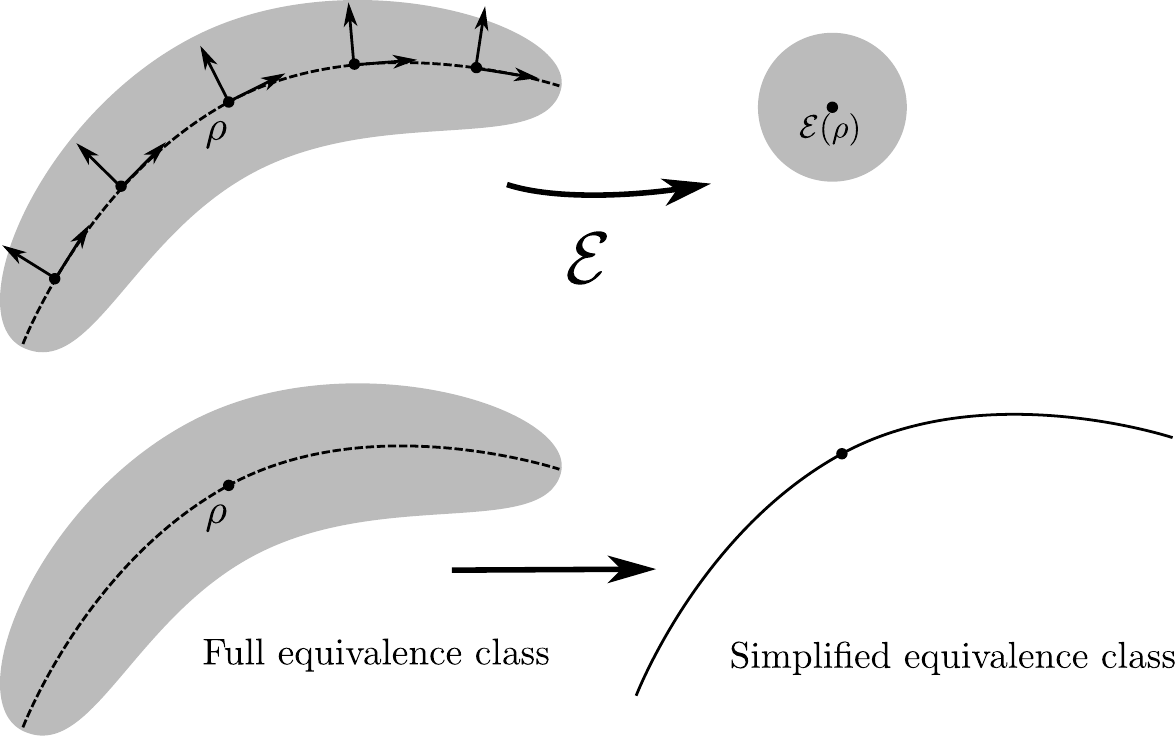} \\
\caption{Nonperturbative version of the approximate equivalence classes of Alice's states: the preimage of the $\epsilon$-ball may be more complex than an ellipsoid due to the nonlinearity of the distinguishability metric. Nonetheless it should be mostly flat along the relevant directions (perpendicular to the integral of the relevant directions passing through $\rho$).
}
\label{ballnonpert} 
\end{figure}

The eigenrelevance operators can be computed for any state $\rho'$. In a finite neighbourhood of a generic state $\rho$ the state-dependant eigenrelevance operators can be chosen \footnote{Actually, we need to choose a \emph{connection} in order to make this identification. We'll elide this point for the moment.} to form continuous tangent fields $X_i(\rho')$, ordered by decreasing eigenrelevance $\eta_j$ at $\rho' = \rho$. Suppose that $\eta_n$ is Bob's chosen relevance threshold. It is reasonable to define the \emph{nonperturbative} equivalence classes of states as submanifolds which are everywhere tangent to the irrelevant fields (Fig.~\ref{ballnonpert}).

\begin{figure}
\includegraphics[width=0.5\columnwidth]{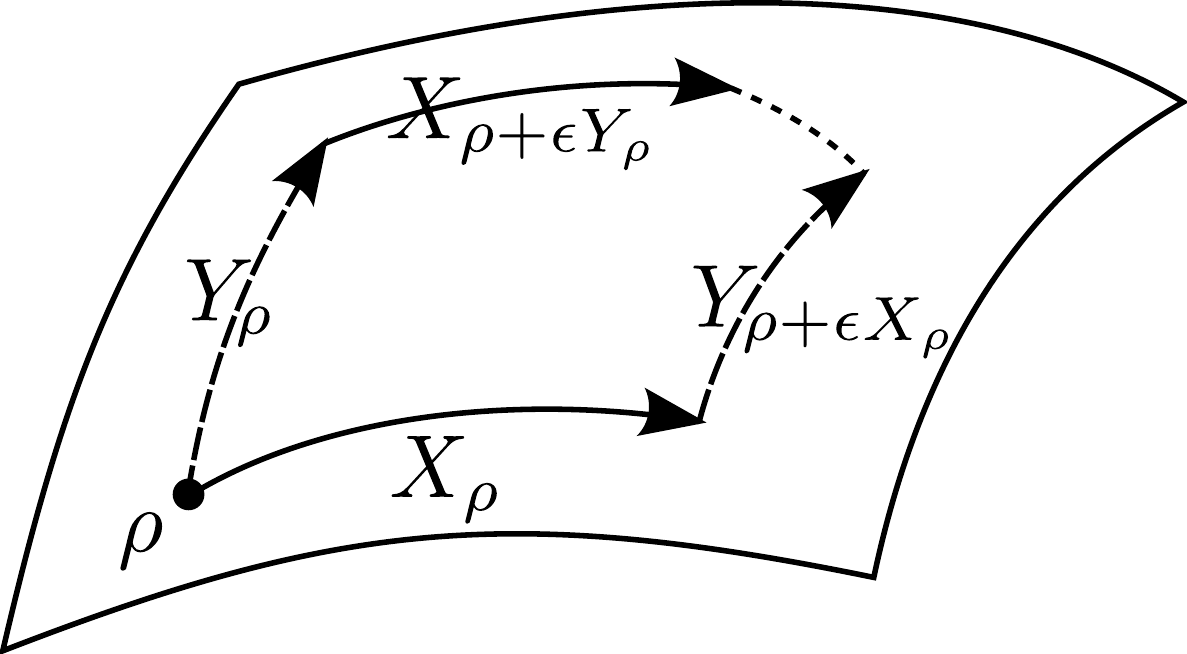} 
\caption{The difference between two infinitesimal paths on a surface must be tangent to the surface.}
\label{frobenius}
\end{figure}  

However, such a foliation does not always exist. The Frobenius theorem of differential geometry states \cite{abraham:1978a} that such a foliation exists if and only if the Lie algebra formed by the irrelevant fields is closed, i.e.,
\[
[X_i,X_j]_\rho = \sum_{k > n} \gamma_{ij}^k X_k(\rho)
\]
for some real numbers $\gamma_{ij}^k$, where $[\cdot,\cdot]$ is the commutator of tangent fields (Fig.~\ref{frobenius}).

If the relevant fields form a closed Lie algebra, then they can be integrated starting from $\rho$. This yields a valid effective manifold, which is everywhere orthogonal to the irrelevant manifolds. We show below that the set of Gaussian states emerge in precisely this way if $\mc E$ is a Gaussian channel.

Apart from this Gaussian example, we do not analyse here the conditions on $\mc E$ so that the vector fields $X_j$ are integrable in the a neighbourhood of a given state, and leave it for future work. Below, we focus on the equivalence conditions derived from the first order analysis.

\section{Examples}

\subsection{Toy model (classical particle)}
\label{toy}

\begin{figure}
\begin{tabularx}{0.9\columnwidth}{XllX}
\includegraphics[width=0.4\columnwidth]{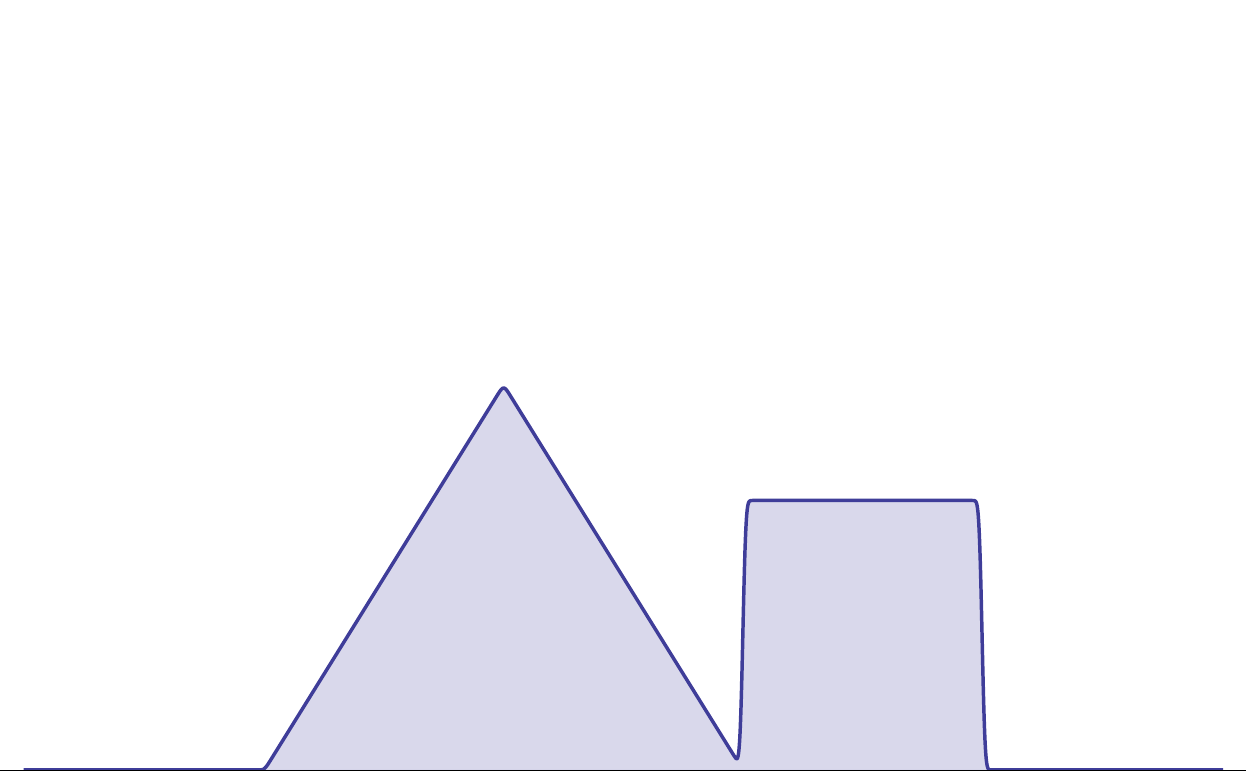} &
\includegraphics[width=0.4\columnwidth]{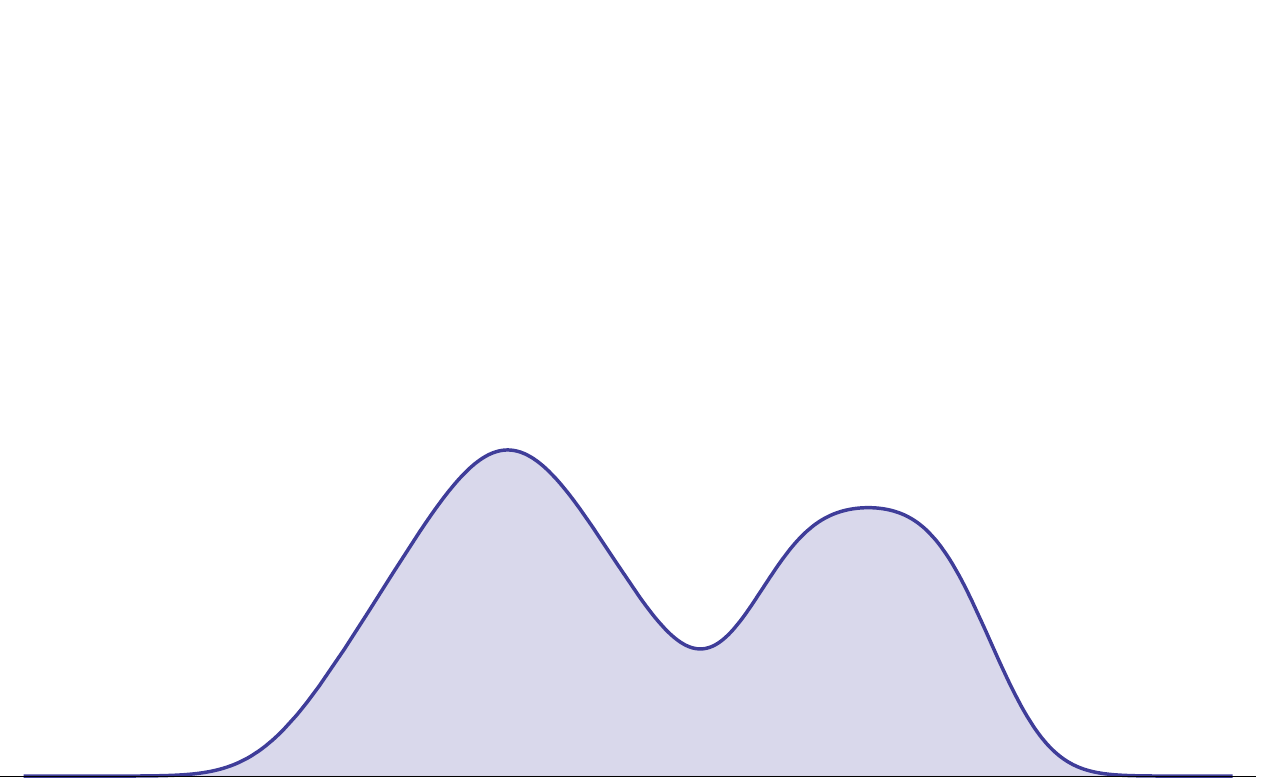} \\
\end{tabularx}
\caption{
Convolution with a Gaussian formalises a limited precision in Bob's measurements of a random variable. 
}
\label{toymodel}
\end{figure}

We first apply our framework to an elementary classical system comprised of a single classical particle in one dimension. This is already enough to illustrate some nontrivial aspects of renormalisation.

In this example the states of both Alice and Bob are probability distributions $x \mapsto \rho(x)$ over $\mathbb R$, and the channel $\mc E$ is the stochastic map given by convolution with a Gaussian
\[
\mc E(\rho)(x) = \frac{1}{\sqrt{2 \pi} \sigma}\int \rho(y) \,e^{-\frac{1}{2\sigma^2}(x-y)^2} dy.
\]
This formalises the idea that Bob can only measure the value of the real number $x$ with precision $\sigma$ (illustrated in Fig.~\ref{toymodel}).

\begin{figure}
\begin{tabularx}{1\columnwidth}{XllX}
\includegraphics[width=0.48\columnwidth]{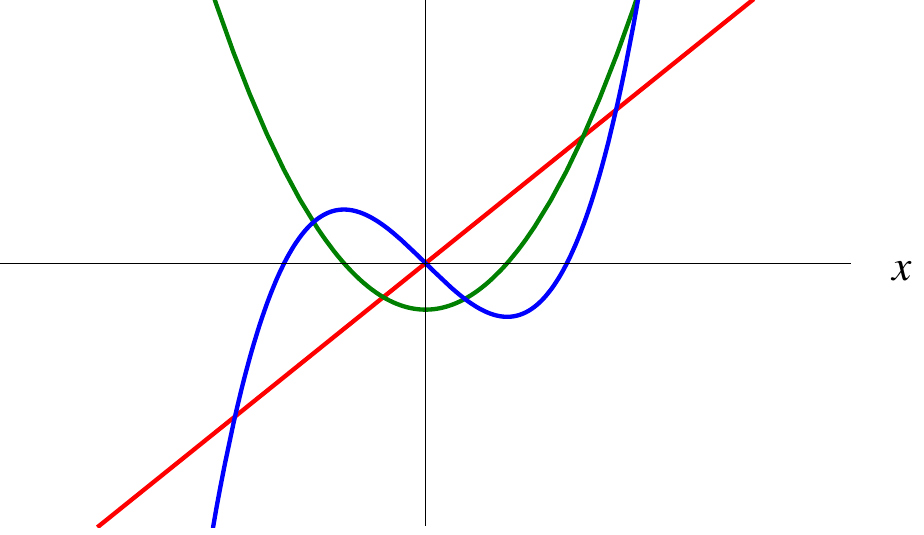} &
\includegraphics[width=0.48\columnwidth]{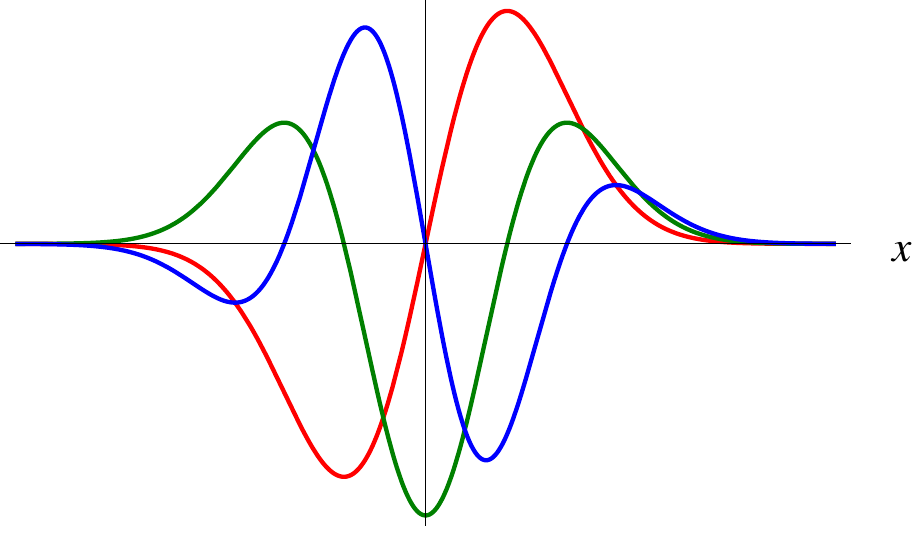} \\
\end{tabularx}
\caption{Toy model of section~\ref{toy}. Left: eigenrelevance observables $A_1$ (red), $A_2$ (green), $A_3$ (blue). Right: corresponding tangent vectors $X_n = \Omega_\rho(A_n) = \rho A_n$. }
\label{toyobs}
\end{figure}

We compute the eigenrelevance directions around a Gaussian state:
\[
\rho(x) \propto e^{-\frac{1}{2 \tau^2} x^2}.
\]
Defining $\alpha = (\sigma^2 + \tau^2)/\tau^2$, we can directly compute
\[
\left({\mc E^\dagger \mc R_\rho^\dagger(A)}\right)(x) = \frac{\alpha}{\sqrt{2 \pi(\alpha^2-1)}\tau} \int A(y) \,e^{-\frac{(x-\alpha y)^2}{2\tau^2(\alpha^2 - 1)}} dy.
\]
The eigenvectors are the Hermite polynomials (Fig.~\ref{toyobs})
\[
A_n(x) = \frac 1 {\sqrt{n!}}{\rm H}_n(x / \tau) = (-\tau)^n  \frac 1 {\sqrt{n!}} e^{\frac{x^2}{2\tau^2}} \frac{d^n}{dx^n} e^{-\frac{x^2}{2\tau^2}},
\]
with eigenvalues 
\[
\eta_n = 1/\alpha^n.
\]
This can be shown using the generating functional $f_t(x) = \sum_n A_n(x) \,t^n/n! = e^{x t/\tau - t^2/2}$ and noticing that $\mc E^\dagger \mc R_\rho^\dagger(f_t) = f_{t/\alpha}$. Comparing the terms of the power series expansion in $t$ on both sides of this equality yields the eigenvectors and their eigenvalues. 

Since the polynomials $A_m$ for $m\le n$ span the polynomials of degree $n$, we can summarise this result by saying that the polynomials of degree $n$ have relevance ratio larger or equal to
\(
\eta_n = \frac{1}{\alpha^n}.
\)
This implies that, if Bob can only accurately measure the $n$ most relevant parameters, then, to first order, he must deem two states to be equivalent if and only if their first $n$ moments are equal. 

As a tangent vector, $A_1$ also generates a change in the distribution expectation value: $-\frac{x^2}{2\tau^2} + \epsilon A_1(x) = -\frac{(x-\epsilon)^2}{2 \tau^2} + \mc O(\epsilon^2)$, and $A_2$ generates a change in the second moment $\tau$: -$\frac {x^2} {2 \tau^2}  + \epsilon A_2(x) = -\frac {x^2} {2 (\tau + \epsilon \tau/\sqrt 2)^2}+ {\rm const.} + \mc O(\epsilon^2)$.  
Since a Gaussian is sent to a Gaussian whenever we move along the two most relevant directions, this shows that the set of all Gaussians $\rho(x) \propto e^{-\frac 1 {2 \tau^2} (x-x_0)^2}$ forms a complete relevant two-dimensional manifold of states. If the irrelevant fields are integrable, then this manifold intersects all the resulting nonperturbative irrelevant manifolds orthogonally. 

Let us use this simple example to see a few ways in which Bob's attempt to determine Alice's state may go wrong.
We assume that Bob chooses to use as effective manifold the exponential family generated by the $n$ most relevant observables:
\[
\rho'(x) \;\propto\; e^{-\frac{1}{2} \frac{x^2}{\tau^2} - \sum_{k=1}^n a_k A_k(x)} = e^{-\frac{1}{2}  \frac{x^2}{\tau^2} - \sum_{k=1}^n b_k (x/\tau)^k}.
\]
The component $a_k$ of a perturbation $B(x) = \sum_{k=1}^n b_k (x/\tau)^k$ is 
\[
a_k = \tr(B\, \Omega_\rho( A_k )).
\]
 
 Suppose Alice's state is anything, but not a Gaussian. As Bob could only determine the two most relevant parameters at first, he was perfectly satisfied with a Gaussian theory $\rho(x) \propto e^{-\frac {x^2} {2 \tau^2}}$, where we use $x_0 = 0$ without loss of generality. His experimentally determined effective Hamiltonian is $H(x) = \frac {x^2} {2 \tau_{\rm phys}^2}$.

However, as he gathers more data, he may be able to attempt to determine higher order terms, such as a fourth order term $x^4$. In Bob's mind, the reason that this term is hard to detect may be that the parameter in front of it is ``small'' (compared to $\tau_{\rm phys}$, his only parameter with a unit). From that point of view, it makes sense to postulate the Hamiltonian $H' = \frac 1 {2 \tau^2} x^2 + \lambda (x/\tau)^4$ with $\tau = \tau_{\rm phys}$. However we know that, in fact, perturbations generated by $A_4$ may be hard to measure for Bob even if $\lambda$ is not small, depending on the value of $\sigma$ and on the number of experiments performed by Bob. 

Of course, because the second moment of the state generated by $H'$ depends on $\lambda$, it is not equal to the parameter $\tau$ entering the fourth-order Hamiltonian, but instead to $\tau_{\rm phys} = (1-6 \lambda) \tau$. Therefore, even before Bob attempts to determine $\lambda$ experimentally, he should at least makes sure that $H'$ is compatible with the old measurements, i.e., it should have the same first two moments as $H$. This is solved by inverting the relationship between $\tau_{\rm phys}$ and $\tau$ and using the parameter $\tau = \tau_{\rm phys} (1 + 6 \lambda)$ in $H'$. In quantum field theory, as shown below, the coefficient in front of $\lambda$ may even be arbitrarily large, making the difference detectable no matter how small $\lambda$ is and how imprecise Bob's measurements are.

Hence Bob has now two effective theories: the more precise one with Hamiltonian $x^2/2\tau^2 + \lambda (x/\tau)^4$, and the less precise $x^2/2\tau_{\rm phys}^2$ which both agree ``at large scale'', i.e., for measurements which are too imprecise to discriminate changes in the state with relevance ratio smaller than $1/\alpha^2$. 

\begin{figure}
\includegraphics[width=0.4\columnwidth]{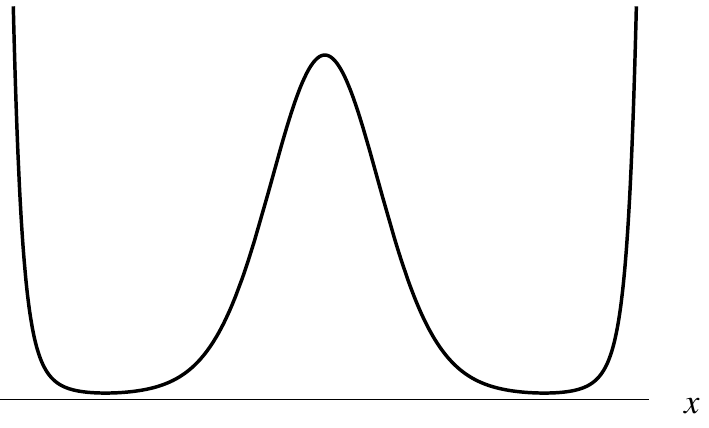} 
\caption{The function $x \mapsto e^{-x^2/2\tau^2 - \lambda (x/\tau)^4}$ cannot define a normalized probability distribution if $\lambda < 0$, even though it does when expended to any order in $\lambda$. }
\label{fig:divergence}
\end{figure} 

Furthermore, for most choices of Alice's true state, a small non-zero value of $\lambda$ will indeed improve Bob's predictions, provided he computes them to first order in $\lambda$. However, if Bob attempts to take this term seriously as a nonperturbative level, he is in for some trouble. Indeed, it may perfectly well be the case that he finds $\lambda < 0$, in which case the resulting state blows up away from the origin and cannot be normalised, leading to {\em infinities} (Fig.~\ref{fig:divergence}). This is somewhat different from the mechanism in which infinities appear in QFT, but it serves our illustrative purpose. 

These infinities can be {\em regularised} by adding a non-zero term proportional to $A_6$, without changing the predictions, yielding a nonperturbatively sound theory. The value of the parameter in front of $A_6$ cannot be determined by Bob because it is beyond his experimental abilities.

Suppose that Bob doesn't know about the eigenrelevance polynomial $A_6$ and instead adds a term of the form $\epsilon (x/\tau)^6$ because for him it seems simpler. Since, unlike $A_6$, the observable $x \mapsto x^6$ has some relevant components, a change in the value of $\epsilon$ would also change the measurable predictions of the theory. Hence, in order to stay within a given experimentally equivalent class, the parameters $m$ and $\tau$ must {\em run} with $\epsilon$ so as to keep the first four moments independant of $\epsilon$.

To first order, the functions $\tau(\epsilon)$ and $\lambda(\epsilon)$ can be simply determined by required that the projection of the Hamiltonian perturbation on $A_2$ and $A_4$ be independant of $\epsilon$. These two components then label the equivalence class on which the curve $\epsilon \mapsto (\tau(\epsilon),\lambda(\epsilon))$ runs.
This leaves open the cosmetic problem of finding a physically more meaningful way of labelling the equivalence class. A possibility is to use the second moment, which we still call $\tau_{\rm phys}$, as well as $\lambda_{\rm phys} := \lambda(0)$. 
In terms of these constants we obtain, to first order, that the bare coupling constants must run as $\lambda(\epsilon) = \lambda_{\rm phys} - 15\, \epsilon$ and $\tau(\epsilon) = {\tau_{\rm phys}}(1 + 6 \, \lambda_{\rm phys} -45 \,\epsilon)$.

\subsection{Classical fields}
\label{classicalfield}

\begin{figure}
\begin{tabularx}{1\columnwidth}{XllX}
\includegraphics[width=0.48\columnwidth]{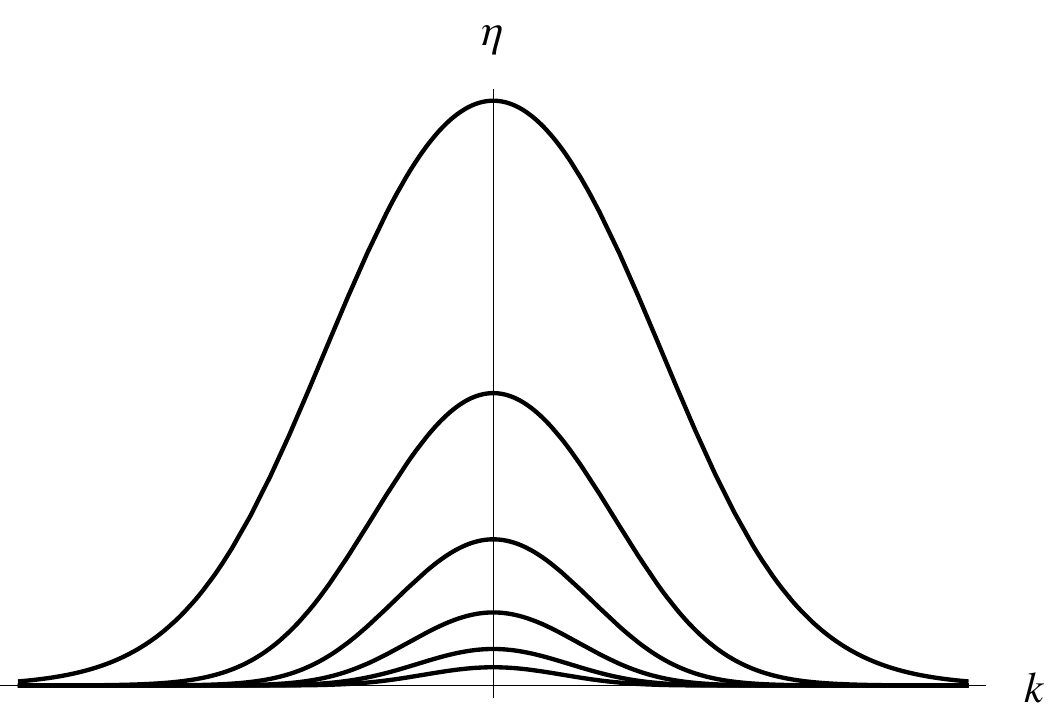} &
\includegraphics[width=0.48\columnwidth]{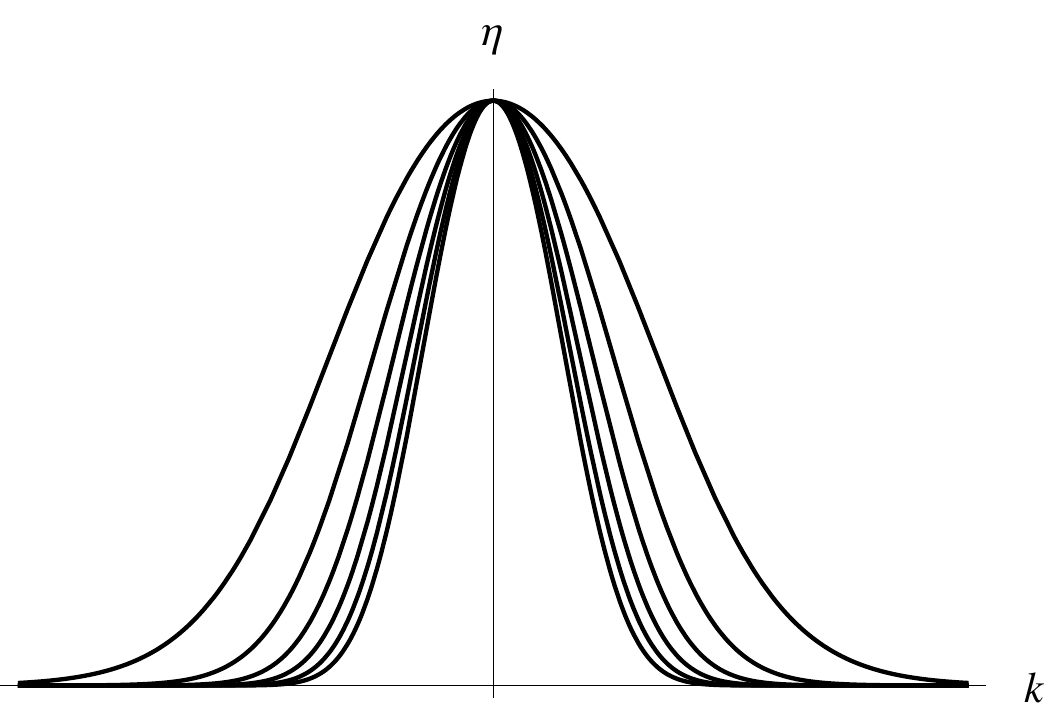} \\
\end{tabularx}
\caption{Relevance eigenvalues $\eta_{k,n}^1$ (Eq.~(\ref{fieldeta})) for the classical scalar field theory with mass (left) and without mass (right). Different curves correspond to different powers $n$. Relevance is larger for larger power $n$ in the field.
The width of the bumps is proportional to the spatial precision $\sigma$ and their vertical separation is governed by the field-value uncertainty $h$. }
\label{fieldrelevances}
\end{figure}

This analysis can be easily extended to classical field theories around a Gaussian state.
We consider real fields $\phi(x)$ in a $d$-dimensional space $x \in \mathbb R^d$. A state is a probability distribution $\rho(\psi)$ over such fields. 
A Gaussian state is of the form
\[
\rho(\phi) \;\propto\; e^{-\frac 1 2 (\phi - \phi_0, A (\phi - \phi_0))}
\]
where the scalar product is the $L^2(\mathbb R^d)$ one, $A$ is an invertible positive linear operator on a suitably defined subset of fields (the covariance operator). In the following we use $\phi_0 = 0$, as this can be easily arranged in any equation by substituting $\phi$ for $\phi - \phi_0$.

It must be noted that this state cannot be naively normalised. Instead, we think of the field formalism as a shorthand for functions on a \emph{finite}, yet arbitrarily large, number of lattice points $x_i \in \mathbb R^d$, and the scalar product is just $(\phi,\psi) = \sum_i \epsilon\, {\phi(x_i)}\psi(x_i)$, where $\epsilon$ is the lattice spacing.

We consider the Gaussian channel (stochastic map) $\mc E$ defined by
\[
\mc E(\rho)(\phi) = \frac{1}{(2\pi h^2)^{N d/2}}\int D\psi \rho(\psi) \,e^{-\frac 1 {2 h^2} (\phi - X \psi, \phi - X \psi)},
\]
where $D\psi \equiv \prod_i d\psi(x_i)$ and  $X$ is an operator with kernel
\[
X(x,y) = N(\sigma) e^{-\frac{1}{2\sigma^2}(x-y)^2}.
\]
This gives the same spatial smudging as the convolution map in the previous example. If the effect of $\mc E$ is interpreted as taking averages of regions of size $\sigma$, then we want to use $N(\sigma) = (2 \pi \sigma^2)^{-d/2}$. The channel has two parameters: $\sigma$ determines the observer's precision in resolving distances, and $h$ his precision in resolving field values. 

Let's consider the case where $A$ and $X$ commute, which happens automatically if we assume that the original Hamiltonian is translation invariant because $X$ is diagonalised by plane waves. We can then label the eigenvectors of $A$ and $X$ by a wavenumber $k$. Let $a_k$ denote the eigenvalue of $A$ for wave number $k$. 
In this plane-wave basis, the modes decouple, and we are left, for each mode, with an instance of the previous one-particle toy model, where $a_k$ plays the role of $1/\tau^2$ an the eigenvalues of $h^2X^{-2}$ play the role of $\sigma^2$.

It follows that the normalised eigenstates of $\mc E^\dagger \mc R^\dagger_\rho$ are 
\[
f^m_{\mathbf k, \mathbf n}(\phi) = \prod_{i=1}^m \frac 1 {\sqrt{n_i!}} {\rm H}_{n_i}( \sqrt{a_{k_i}}\, \phi_{k_i}).
\]
where
\[
\phi_k := \int dx \,\phi(x)\, \cos(i k x). 
\]
They are labelled by an integer $m$, a choice of $m$ distinct modes $\mathbf k = (k_1,\dots,k_m)$, and a choice of integer degree for each mode: $\mathbf n = (n_1,\dots,n_m)$.  
The corresponding eigenvalues (relevance ratios) are
\begin{equation}
\label{fieldeta}
\eta_{\mathbf k, \mathbf n}^m = \prod_{i=1}^m \left({ 1 + a_{k_i} h^2 e^{k_i^2 \sigma^2}  }\right)^{-n_i}.
\end{equation}

The exponential factor in Eq.~(\ref{fieldeta}) effectively renders any mode with $k > 1/\sigma$ irrelevant. Hence the spatial precision parameter $\sigma$ acts as a momentum cutoff. However, the relevance of low momentum modes depends on the power of the field operators only through the parameter $h$ which characterises the observer's precision in measuring field values.

As an example, we consider the thermal state for a massive classical scalar field, with
\[
a_k = \beta \sum_i k_i^2 + m^2.
\]

In particular, if $m>0$ and we keep only modes with $k \ll 1/\sigma$ then the relevance of the quadratic polynomials in the fields asymptotically separates from that of higher order polynomials as $\sigma \rightarrow \infty$. Since the translation-invariant quadratic observables are tangent to the manifold of Gaussian states, we see that this manifold forms a good relevant nonperturbative effective manifold for translation-invariant theories. Notice that for $m=0$, however, all powers of the fields at $k=0$ are equally relevant. This is a sign of criticality: any long wavelength perturbation around the state can be easily detected by the observer. 

We defer the discussion of the renormalisation group in this model to the quantum case below. 

Apart for $k=0$, none of the eigenrelevance observables are translation invariant. The relevance of a translation-invariant operator can be computed by finding its components in terms of the eigenrelevance observables. For instance, consider $A(\phi) = \int dx \,\phi(x)^2$ (and taking $\phi_0 = 0$ for simplicity). It can be written as
\[
A(\phi) = \sum_{k} \,\phi_k^2 = \sum_k \sqrt 2 \,a_k^{-1} f^1_{k,2}(\phi) + \sum_k a_k^{-1}.
\]
Once we subtract the non-trace-preserving constant term $A_0 = \sum_k a_k^{-1}$, the tangent vector $A - A_0$ has relevance
\[
\eta(A) = \frac{\sum_k  a_k^{-2} \eta^1_{k,2}}{\sum_k a_k^{-2}}.
\]
The sum in the numerator is effectively cutoff at $k \approx 1/\sigma$ because of the relevance parameter, and is therefore finite even in the continuum limit. However, the sum in the denominator diverges and requires a finite lattice spacing $\epsilon$, or ultraviolet (UV) cutoff. 

Asymptotically, for $\sigma \gg 1/m$, $\eta(A-A_0)$ behaves in terms of $\sigma$ and $h$ like $\mc O(\sigma^{-d} h^{-4})$, where $d$ is the dimension of space. This can be compared to the perturbation $B(\phi) = \int dx \,\phi(x) \partial_i \partial^i \phi(x)$, whose relevance scales as
\(
\eta(B-B_0) = \mc O(\sigma^{-d-2} h^{-4}).
\)
Hence, the observable $B$ becomes harder to measure compared to $A$ as Bob becomes less accurate in his spatial measurements. This matches the RG idea that the Hamiltonian $H = B$ is an unstable ``fixed point'', while $H = A$ is stable. However, no parameter is obviously flowing in this picture and we cannot simply drop the less relevant term $B$ in the Hamiltonian because it is not orthogonal to $A$. Below, we show how to derive a proper renormalisaton flow as a function of $\sigma$ in the quantum case by dropping eigenrelevant terms.

\subsection{Quantum particle}
Here we discuss the eigenrelevant operators for a single quantum particle moving in one dimension with canonical observables $\hat x$ and $\hat p$. The hypothesis $\rho$ for Alice's state, in this case, is taken to be a Gaussian quantum state.

A Gaussian state $\rho$ with characteristic function
\[
\chi_\rho(x,p) = e^{-\frac 1 4 (u^2 x^2 + v^2 p^2) + p_0 x - x_0 p},
\] 
where $u,v$ are positive and $uv \ge 1$,
can be written as $\rho = e^{-H}$, where
\begin{equation}
\label{qgh}
H = \coth^{-1}(u v) \left({ \frac u v (\hat x - x_0)^2 + \frac v u (\hat p - p_0)^2 }\right) + \alpha \one
\end{equation}
and 
\[
\alpha = \log\sqrt{u^2 v^2 - 1}.
\]

A lack of precision in measuring the position $\hat x$ and momentum $\hat p$ (or field observables if this is a mode) can be formalised as a Gaussian channel which maps $u^2$ to $u^2 + \sigma_p^2$ and $v^2$ to $v^2 + \sigma_x^2$,
where $\sigma_x$ and $\sigma_p$ are the uncertainties is measuring $\hat x$ and $\hat p$ respectively.
This corresponds to taking a linear combination of Gaussian displacements of the particle in position and momentum. 

Knowing that $\Omega_\rho$ is the operator derivative of the exponential function, and $\Omega_\rho^{-1}$ the derivative of the logarithm, it is easy to see that, in general,
\begin{equation}
\label{quantgauss}
\mc E(e^{-H + \epsilon A}) = e^{-H' + \epsilon\, \mc R_\rho^\dagger(A)} + \mc O(\epsilon^2),
\end{equation}
where $\rho \propto e^{-H}$ and $\mc E(\rho) \propto e^{-H'}$. 

This implies that a quadratic $A$ is mapped to a quadratic $\mc R_\rho^\dagger (A)$. Since $\mc E^\dagger$ also maps quadratic terms to quadratic terms the two eigenvectors of $\mc E^\dagger \mc R_\rho^\dagger$ must be second order polynomials in $\hat x$ and $\hat p$.

We find that both $\hat x$ and $\hat p$ are eigenvectors.
Asymptotically for large $\sigma_x$ and $\sigma_p$, their relevances are
\[
\eta(\hat x) \approx \frac v {s u} \,\sigma_x^{-2}\quad \text{and} \quad\eta(\hat p) \approx \frac{u}{s v} \sigma_p^{-2},
\]
where we used $s = \coth^{-1}(uv)$.

In terms of $u$ and $v$, the second order eigenvectors are complicated linear combinations of $\hat x^2$, $\hat p^2$ and $\one$, even asymptotically for large $\sigma_x$ and $\sigma_p$. However, if the state $\rho$ is very mixed ($uv \rightarrow \infty$), then we find the eigenvectors $\hat x^2 - \frac s 2 \frac{u}{v}\one$ and $\hat p^2 - \frac s 2 \frac{v}{u}\one$ with respective eigenvalues $u^4 \sigma_p^{-4}$ and $v^4 \sigma_x^{-4}$.

\subsection{Quantum fields}
\label{quantfields}

In general, a quantum Gaussian channel is defined by two real matrices $X$ and $Y$, such that its effect on a Gaussian state's covariance matrix $\gamma$ is
\[
\gamma \mapsto X^T \gamma X + Y.
\]
These operators are not independent, as they must satisfy $Y + iS - X^\dagger iS X \ge 0$, where $S$ is the kernel of the symplectic inner product.  Simultaneously, the expected field $\phi_0$, if nonzero, is mapped to $X \phi_0$.

In order to define Bob's lack of spatial precision, one may use the same spatial mode mixing operator $X$ parametrized by $\sigma$ as in the classical case, assuming it acts identically on the position and the momentum degrees of freedom. A lack of precision in measuring field values can be simulated by a matrix $Y$ which is proportional to the identity on the field coordinates and on the field canonical conjugates, but with different coefficients. 
In the neighbourhood of a translation-invariant quadratic theory the effect of this channel factors for each momentum mode as in the classical field example.

For concreteness, we consider a scalar field theory. The Hamiltonian is 
\[
H = \frac 1 2 \int dk \,(\Pi_k^2 + \omega_k^2 \Phi_k^2),
\]
where $\omega_k = \sqrt{k^2 + m^2}$ and, in terms of the Fourier transforms $\phi_k$ and $\pi_k$ of the canonical field operators $\phi(x)$ and $\pi(x)$,  
\[
\Phi_k = \re \phi_k - \frac 1 \omega_k \im \pi_k\quad\text{and}\quad\Pi_k = \re \pi_k + \omega_k \im \phi_k.
\]

The effect of the channel $\mc E$ on states of the form
\[
\rho \propto e^{- \int dk  \coth^{-1}(u_k v_k) \left({\frac{v_k}{u_k}(\Pi_k-\delta_k \one)^2 + \frac{u_k}{v_k} (\Phi_k-\epsilon_k \one)^2}\right)},
\]
is to map $u_k^2$ to $X_k^2 u_k^2 + 2h_\Phi^2$ and $v_k^2$ to $X_k^2 v_k^2 + 2h_\Pi^2$, and $\delta_k$ to $X_k \delta_k$ and $\epsilon_k$ to $X_k \epsilon_k$,
where $X_k = e^{-\frac 1 2 k^2 \sigma^2}$, and $h_\Phi$ and $h_\Phi$ parameterize the precision at which the fields are resolved. 

By using the state 
\[
\rho \propto e^{- \frac \beta 2 \int dk \left({(\Pi_k-\delta_k \one)^2 + \omega_k^2 (\Phi_k-\epsilon_k \one)^2}\right)}
\]
and looking at the linear terms in $\epsilon_k$ and $\delta_k$ using Eq.~(\ref{quantgauss}), we deduce the effect of $\mc R_\rho^\dagger$ on $\Pi_k$ and $\Phi_k$. Combined with the fact that $\mc E^\dagger(\Pi_k) = X_k \Pi_k$ and $\mc E^\dagger(\Phi_k) = X_k \Phi_k$, we obtain, asymptotically for $h_\Phi h_\Pi \gg 1$, the eigen-relevances
\[
\eta(\Phi_k) \simeq \frac 1 {\frac{\beta \omega_k} 2 \coth\frac{\beta\omega_k}2 + \beta \omega_k^2 h_\Phi^{2} \, e^{k^2\sigma^2}}
\]
and
\[
\eta(\Pi_k)  \simeq \frac 1 {\frac{\beta \omega_k} 2 \coth \frac{\beta\omega_k}2 + \beta h_\Phi^{2} \,  e^{k^2\sigma^2}}.
\]

Since the channel acts independently on each mode, then the products $\Phi_{k_1} \cdots \Phi_{k_n}$, for instance, are eigen-relevant with relevance $\eta(\Phi_{k_1}) \cdots \eta(\Phi_{k_n})$, provided that the momenta $k_1,\dots,k_n$ are all distinct. 

Recall that our first-order prescription says that two effective Hamiltonians are effectively equivalent if they yield the same expectation values for eigenrelevance observables down to the chosen minimal relevance level. In this case, the eigenrelevance observables are the $n$-point correlation functions, with relevance decreasing exponentially with $n$ (for momenta $k \ll 1/\sigma$). 

This is precisely how the renormalisation conditions are derived in standard quantum field theory: by running the coupling constant with the cutoff in such manner that the $n$-point correlation functions stay constant. Typical effective Hamiltonians are such that, indeed, only the first few $n$ are needed to fix all the parameters.

\subsection{Wilsonian renormalisation}
\label{sec:wilson}

From the above analysis, we also obtain that two independent linear combinations of $\Phi_k^2$, $\Pi_k^2$ and $\one$  are eigenrelevant and that, to leading order in $\sigma$, their relevance 
decreases exponentially with $k$ as $e^{-k^2 \sigma^2}$. 
This implies that the Hamiltonian $H_{\epsilon} = \frac 1 2 \int_{|k| < 1/\epsilon} dk \,(\Pi_k^2 + \omega_k^2 \Phi_k^2)$, with regularisation parameter $\epsilon$, is in the same equivalence class as
\begin{equation}
\label{cutoffham}
H_{\sigma} = \frac 1 2 \int_{|k| < 1/\sigma} dk \,(\Pi_k^2 + (k^2 + m^2) \Phi_k^2).
\end{equation}

To first order, this holds because the difference $H_\sigma - H_\epsilon$ consists only of eigen-relevant terms of small enough relevance. But this also holds to all orders in the expension of the exponential $e^{-\beta H}$ (albeit using the first-order definition of the equivalence classes) due to that fact that the high and low momentum terms are decoupled, and hence removing the high momentum terms does not influence the $n$-point correlation functions for modes $k < 1/\sigma$. 

This means that we can simply drop the irrelevant high-momentum quadratic terms to simplify the Hamiltonian as the imprecision $\sigma$ increases. However, if the Hamiltonian also contains the term $\lambda \int dx \phi^4(x) = \lambda \int dk_1 \cdots dk_4 \phi_{k_1}\cdots \phi_{k_4} \delta(k_1 + \dots + k_4)$, for instance, then simply changing the bound of the momentum integrals from $\epsilon$ to $\sigma$ would put the state in a different equivalent class, unless the parameters $m$ and $\lambda$ are modified as a function of $\sigma$ so as to preserve the $n$-point correlation functions for modes $k < 1/\sigma$. 

This procedure defines a continuous renormalisation flow in terms of $\sigma$; mathematically, it is also precisely the one we would use to determine the change in the Hamiltonian's parameters needed to compensate for a change in the regularisation parameter from $\epsilon$ to $\sigma$, in order to stay within the same equivalence class. Hence the two completely different types of renormalisation flow mentioned in the introduction---in terms of the precision parameter $\sigma$ or in terms of the regularisation parameter $\epsilon$---happen to be identical in this example. 

We have not yet mentioned the role of {\em scaling} which is prevalent in Wilson's approach to renormalisation. 
We saw that an increase in the precision parameter $\sigma$, and the subsequent discarding of newly irrelevant terms, manifests itself in two very different ways: a change of momentum cutoff in the Hamiltonian, as well as a possible change of the ``coupling constants'', i.e., parameters in the integrand. However, the change of cutoff can also be treated as a change in the coupling constants by simply {\em rescaling} space.
Indeed, the Hamiltonian in Eq.~(\ref{cutoffham}) can also be rewritten with the same cutoff $1/\epsilon$ as before via a change of variable corresponding to a scaling transformation 
\(
\tilde k = k/s,
\)
\[
\tilde \Phi_{\tilde k} = s^{\frac{d+1}2} \Phi_{s \tilde k} 
\quad \text{and} \quad
\tilde \Pi_{\tilde k} = s^{\frac{d-1}2} \Phi_{s \tilde k}, 
\]
where $s = \epsilon/\sigma$, so that we can write
\[
H_\sigma = \frac s 2 \int_{|\tilde k| < 1/\epsilon} d\tilde k \,\left[{ \tilde \Pi_k^2 + (\tilde k^2 + s^{-2} m^2) \tilde \Phi_{\tilde k}^2}\right].
\]
The factor $s$ in front of the Hamiltonian is compensated by also scaling the temperature as $\tilde \beta = s \beta$ (which can be thought of as imaginary time, hence scaling like a spatial coordinate).

This shows that removing the high momentum terms in the Hamiltonian is equivalent to scaling the system up (and hence also the cutoff) while increasing the mass to $\tilde m = s^{-2} m^2$. Any term in the Hamiltonian would take in this way a trivial dependance on $\sigma$ mirroring the neglect of high momentum terms in addition to its possibly non-trivial dependence needed to keep the state in the same equivalence class.

\subsection{Momentum shell RG}

In the previous section, we partly neglected the effect of the field value imprecision on the relevance of observables. This approximation can also be performed earlier in our analysis. 

If we make $\sigma$ very large while keeping $h_\Phi$ and $h_\Pi$ fixed, the effect of $\mc E$ may be idealized by a channel $\mc E_\sigma$ which simply traces out all momentum modes with wave-vector of norm larger than $1/\sigma$. If the state $\rho$ factors in terms of these modes, which is the case if $\rho$ is Gaussian and translation invariant, then $\mc E^\dagger \mc R_\rho^\dagger$ is simply a projector on the space of operators acting trivially on modes with wave vectors larger than $1/\sigma$. 

In order to see this, let us write the state as $\rho = \rho_< \otimes \rho_>$ where the first system is that composed of the modes with wave vectors smaller than $1/\sigma$. Then, noting that $\mc E(\rho) = \rho_<$, a direct calculation shows that for all $A\otimes B$, $\mc E^\dagger \mc R_\rho^\dagger(A\otimes B) = A \otimes \one \,\tr(\rho_{>} B)$. In particular, this implies that all operators of the form $A \otimes \one$ have eigenvalue one, and all operators of the form $A\otimes B_0$ where $\tr(\rho_{>} B_0)=0$ have eigenvalue zero. Since these span all operators on the joint system, this proves the statement. 

We see that the experimental limitations defined by $\mc E_\sigma$ give us much less guidance on how to define our effective theory in a neighbourhood of $\rho$. For instance, at least to first order, it does not assign different relevance value to different powers of the field operators. 

Nevertheless, for a given family of effective states, it is enough to remove the ambiguities coming from neglecting small scale features. For instance, let's consider again a relativistic free scalar quantum field theory. The Hamiltonian contains the mass term $\frac {m^2} 2 \int dx\, \phi(x)^2$, where $\phi(x)$ is the self-adjoint field operator. It is connected to the annihilation operators $a_k$ through $\phi_k = \frac{1}{\sqrt{2 \omega_k}}(a_k + a_{-k}^\dagger)$, where $\omega_k = \sqrt{|k|^2 + m^2}$.
Also we assume a UV cutoff defined by the minimum length $\epsilon$. Let $\rho$ be its state at some finite temperature. We write $\ave{A} = \tr(\rho\,A)$ for any operator $A$. Suppose we add an interaction term $A = \frac{\lambda}{4!} \int dx\, \phi(x)^4$ to the Hamiltonian. In terms of momentum modes $\phi_k$, this term has the form
\[
A = \frac{\lambda}{4!} \int dk_1\cdots d_{k_4} \phi_{k_1} \cdots \phi_{k_4} \delta(k_1 + \dots + k_4). 
\]
The projection $\mc E^\dagger \mc R_\rho^\dagger(A)$ on the relevance one subspace of operator contains a term $A_2$ of second order in the field. In the zero temperature limit, it is
\[
A_2 = \frac{ \lambda}{4}  \int_0^{1/\sigma} dk' dk'' \int_{1/\sigma}^{1/\epsilon}\frac{dk}{2\omega_k}\,\delta(k'+k'')\phi_{k'} \phi_{k''} 
\]
where we use the fact that $ \ave{\phi_{k} \phi_{-k'}} = \frac{\delta(k+k')}{2\omega_k}$. Also, the bounds on the integral signify upper and lower bounds to the Euclidean norm $|k|= \sum_i k_i^2$ of the spatial wavevector $k$. This reduces to
\[
A_2 = \frac{\lambda}{4}  \int_{1/\sigma}^{1/\epsilon}  \frac{dk'}{2\omega_{k'}}\int_0^{1/\sigma} dk\, \phi_{-k} \phi_{k},
\] 
where $\int_0^{1/\sigma} dk\, \phi_{-k} \phi_{k}$ is, up to a constant, the projection on the relevant manifold of the quadratic term
\(
\int dx \,\phi(x)^2 = \int dk\, \phi_{-k} \phi_{k}.
\)
Hence, to first order in $\lambda$, the physical mass is 
\begin{equation}
\label{regul}
m_{\rm phys}^2 = m^2 +  \frac{\lambda}{2}  \int_{1/\sigma}^{1/\epsilon}  \frac{dk}{2\omega_k}.
\end{equation}
In the limit $\sigma \rightarrow \infty$, this matches the usual result from momentum cutoff regularisation, as $\int_{0}^{1/\epsilon} \frac{dk}{2\omega_k}$ is the regularised propagator $G(x) = \ave{\phi(x)\phi(0)}$ at $x=0$. For finite $\sigma$, the result smoothly interpolates down to the case $\sigma = \epsilon$ where Bob has the means to measure the ``bare mass'' directly.



\section{Discussion}

In this paper we have introduced an information-theoretic formulation of the RG, appropriate for both the statistical physics and quantum field settings. We achieved this by first describing a game involving two players, Alice, who has a system and Bob, who can only perceive the system via a lossy quantum channel modelling his experimental limitations. Bob's objective is to infer the state of Alice's system, which is an ill-posed inverse problem. We showed how to render this inverse problem well posed by: (i) working in the neighbourhood of an initial reasonable hypothesis; and (ii) decomposing this neighbourhood into equivalance classes of states determined by the least relevant degrees of freedom.

Each equivalance class is a convenient idealizations of a set of hypothesis about Alice's state which Bob cannot distinguish given his limited information.

An \emph{effective theory} is then a smooth parametrisation of these equivalence classes, such as a submanifold of states which intersects each class at exactly one point. 

The manifestations of the RG in statistical physics and quantum field theory were then described in this new information-theoretic setting: in the statistical physics setting we showed that the RG is associated with a flow on an equivalence class whereby Bob tries to find a \emph{simplification} within the class for his effective theory of Alice's system when he increases a noise parameter. In quantum field theory Bob also obtains a flow on an equivalence class, however, this time the flow is induced by an arbitrary regulator required to keep the system's state well defined. 

Finally we calculated the eigenrelevance observables around a gaussian hypothesis state in a variety of settings from that of a single classical particle to a scalar quantum field theory. Given a reasonable model of Bob's limitations we showed that the manifold of Gaussian states is everywhere tangent to the most relevant directions. In addition, this same model appears to justify the use of $n$-point correlations functions in summarising the predictions of a field theory up to a given level of confidence. We have not, however, completed the characterization of all the eigenrelevance observables around a quantum gaussian mode, which should be feasible.

Interestingly, these results correspond only to a ``first-order'' approximation of the irrelevant manifolds. As one moves further away from Gaussian states, the nature of the irrelevant observables may change because of the nonlinearity of the information metric. Those irrelevant manifolds, or equivalence classes of states that cannot be experimentally distinguished, are not necessarily well defined beyond a first order analysis, as the irrelevant fields are not necessarily integrable. This leaves open the question of what are the conditions on $\mc E$ so that these manifolds are well-defined to higher order, or even nonperturbatively near certain states. Secondly, it is not currently  a priori  clear how important such more precise characterisations of the equivalance classes would be in practice. 

We end with a list of open questions and potential applications of this work.

\begin{enumerate}
\item Although we only analysed the inverse problem in the neighbourhood of Gaussian states, a very interesting aspect of this approach is that it should allow one to extend renormalisation group techniques to widely different choices of effective states. For instance, one could postulate a nongaussian \emph{tensor network state} as a hypothesis for Alice's state and calculate the eigenrelevance observables under the channel given by, e.g., Kadanoff block-spin renormalisation.

\item This formalism could be used to study the classical limit of quantum theory. We see that in the context of Gaussian quantum field theories it provides an operational justification for why an observer would effectively only have access to the expectation values of a limited set of observables (rather than, say, full outcome probabilities). Together with results such as the Ehrenfest theorem, this may provide the justification for the emergence of classical effective models. In fact, the results of section~\ref{quantfields} may justify the use of the effective action, which encodes the expectation values of the field operators. 

\item Another completely different application is to understand the situations where an infinite-dimensional system can be effectively modelled in terms a finite-dimensional Hilbert space, or even just one qubit.

\item In the Gaussian examples studied here the expectation values of the fields appear as the most relevant variables, while their second moments (fluctuations) come as the next most relevant. Can this approach be related to the classical and quantum central limit theorems, where the value of the noise parameter is related to the power of the $1/N$ normalization factor in the front of a sum of $N$ random variables?

\item There are many instances of the RG in condensed matter physics, particularly as numerical methods. It would be interesting to investigate the formulation of such numerical RG methods in terms of our information-theoretic formalism. In particular, we expect that quantifying the eigenrelevance observables in this case may lead to faster numerical methods whereby certain variational degrees of freedom can be consistently neglected.

\item The calculations presented in this paper were only carried out for the metric arising from the relative entropy; in the classical case this is the unique monotone information metric. In the quantum case, however, there are infinitely many monotone riemannian metrics. It would be interesting to understand what effect the change in metric would have in the quantum case. For example, the recently introduced $\chi^2$ divergence \cite{temme:2010a} enjoys an operational interpretation which is arguably closer to some experimental situations.

\item Quantum field theory is understood to be a good effective description of critical models in statistical physics. An intriguing open problem is to see how such continuum limits for quantum systems can arise in our RG framework.\

\item What happens when we replace quantum mechanics with a more general probabilistic theory? Could it be that quantum mechanics itself arises as a good effective theory for Alice's system? Partial evidence for this possibility has recently been discussed in \cite{kleinmann:2013a}.  

\item How about the emergence of thermodynamics as an effective theory? For example, in the case of an experimentalist with a single (imprecise) observable we should get the Boltzmann state as a good effective state. What happens when we add observables?

\item What properties of the family of channels, and of the initial hypothesis, guarantee that if an observable is irrelevant at a given noise level, it stays fully irrelevant for a higher value of the noise parameter?

\item We haven't investigated the role of symmetries in our picture: how does postulating a global or local symmetry simplify the calculation of the eigenrelevance observables?

\item Only the thermal -- imaginary time -- case was considered here. Can the formalism be extended to a situation where the experimentlist attempts to determine the system's dynamics?

\item In the context of classical inference from data, Transtrum et al.~\cite{transtrum10} observed that in many models a hierarchical structure is apparent not just locally, but also in the global dimensions of the manifold of models. Can such results be used to better understand the possible non-perturbative extensions of our framework?

\end{enumerate}

\section{Acknowledgments}
Helpful discussions with numerous people are most gratefully acknowledged: a partial list includes Andrew Doherty, Jens Eisert, Steve Flammia, Jutho Haegeman, Gerard Milburn, Terry Rudolph, Tom Stace, Frank Verstraete, and Reinhard Werner. This work was supported by the ERC grant QFTCMPS and by the cluster of excellence EXC 201 Quantum Engineering and Space-Time Research.

\bibliography{renormalization_primer}

\end{document}